\documentclass[12pt,onecolumn,draftcls]{IEEEtran}
\usepackage{cite}
\usepackage{amsmath,amssymb,amsfonts,acronym}
\usepackage{algorithmic}
\usepackage{textcomp}
\usepackage[usenames]{color}
\usepackage{arydshln}
\usepackage{bm}
\usepackage{url}
\usepackage{soul}
\usepackage{comment}
\usepackage{xcolor,colortbl}
\usepackage{graphicx,color}
\usepackage{pstool}
\definecolor{mycolor}{rgb}{0,0.5,0.5}
\definecolor{orange}{rgb}{1,0.5,0}

\definecolor{LightCyan}{rgb}{0.88,1,1}

\acrodef{BIM}{Bayesian information matrix}

\acrodef{CRLB}{Cram\'er-Rao lower bound}

\acrodef{DOA}{direction-of-arrival}
\acrodef{EKF}{Extended Kalman Filter}
\acrodef{FANET}{flying ad-hoc network} 
\acrodef{FIM}{Fisher Information Matrix}
\acrodef{FMCW}{frequency-modulated continuous-wave}
\acrodef{GPS}{Global Positioning System}
\acrodef{GMTI}{Ground Moving Target Indicator}
\acrodef{HPBW}{Half Power Beamwidth}
\acrodef{LOS}{line-of-sight}
\acrodef{MILP}{Mixed Integer Linear Programming}
\acrodef{mm-wave}{millimeter wave}
\acrodef{MCP}{Model Predictive Control}
\acrodef{MAV}{micro aerial vehicle}
\acrodef{ML}{maximum likelihood}
\acrodef{MSE}{mean squared error}
\acrodef{NLOS}{non line-of-sight}
\acrodef{NFER} {Near-Field Electromagnetic Ranging}
\acrodef{PEB}{Position Error Bound}
\acrodef{RV}{random variable}
\acrodef{RSS}{received signal strength}
\acrodef{RMSE}{root mean squared error}
\acrodef{RCS}{radar cross section}
\acrodef{RTT}{round-trip time}
\acrodef{SLAM}{Simultaneous Localization and Mapping}
\acrodef{std}{standard deviation}
\acrodef{SNR}{signal-to-noise ratio}
\acrodef{TOA}{time of arrival}
\acrodef{UAV}{unmanned aerial vehicle}
\acrodef{UWB}{ultrawide bandwidth}
\acrodef{WSN}{wireless sensor network}
\acrodef{WRN}{wireless radar network}
\acrodef{WSR}{wireless sensor radar}

\acrodef{DRN}{dynamic radar network} 

\acrodef{ATC}{adapt-then-combine} 

\newcommand{\posi} {\mathbf{p}_{i}^{(k)}} 
\newcommand{\fposi} {\mathbf{p}_{i}^{(k+1)}}
\newcommand{\posjl} {\mathbf{p}_j^{(\ell_k)}} 
\newcommand{\kposi} {\mathbf{L}_i^{(k)}} 
\newcommand{\fkposi} {\mathbf{L}_i^{(k+1)}} 
\newcommand{\xik} {x_{i}^{(k)}} 
\newcommand{\yik} {y_{i}^{(k)}} 
\newcommand{\zik} {z_{i}^{(k)}} 
\newcommand{\fxik} {x_{i}^{(k+1)}}
\newcommand{\fyik} {y_{i}^{(k+1)}}
\newcommand{\fzik} {z_{i}^{(k+1)}}
\newcommand{\vvik}{\mathbf{v}_{i}^{(k)}} 
\newcommand{\ui} {\mathbf{u}_{i}^{(k)}} 
\newcommand{\fui} {\mathbf{u}_{i}^{(k+1)}}
\newcommand{\uxi} {u_{\mathsf{x},i}^{(k)}} 
\newcommand{\uyi} {u_{\mathsf{y},i}^{(k)}}
\newcommand{\uzi} {u_{\mathsf{z},i}^{(k)}}
\newcommand{\speedik} {v_{i}^{(k)}} 
\newcommand{\speedmin}{v_{\mathsf{min}}}
\newcommand{\speedmax}{v_{\mathsf{max}}}
\newcommand{\headik} {\Psi_i^{(k)}}

\newcommand{\headmax}{\Psi_{\mathsf{max}}}
\newcommand{\tiltik} {\Theta_i^{(k)}}

\newcommand{\tiltmax}{\Theta_{\mathsf{max}}}
\newcommand{\pik}{\mathrm{l}_i^{(k)}}
\newcommand{\pjl}{\mathrm{l}_j^{(\ell_k)}} 
\newcommand{\vpik}{\bm{\mathrm{l}}_i^{(k)}} 
\newcommand{\posk} {\mathbf{p}_0^{(k)}}

\newcommand{\xok}{x_0^{(k)}}
\newcommand{\yok}{y_0^{(k)}}
\newcommand{\zok}{z_0^{(k)}}
\newcommand{\eposk} {\hat{\mathbf{p}}_{0; i}^{(k)}}
\newcommand{\eeposk} {\hat{\mathbf{p}}_{0; i}^{(k+1 \lvert k)}}
\newcommand{\eposkm} {\hat{\mathbf{p}}_{0; im}^{(k)}}
\newcommand{\velk} {\mathbf{v}_0^{(k)}}

\newcommand{\vox}{v_{\mathsf{x},0}^{(k)}}
\newcommand{\voy}{v_{\mathsf{y},0}^{(k)}}
\newcommand{\voz}{v_{\mathsf{z},0}^{(k)}}
\newcommand{\statek} {\mathbf{s}^{(k)}}
\newcommand{\estatek} {\hat{\mathbf{s}}_{ i}^{(k)}}
\newcommand{\eestatek} {\hat{\mathbf{s}}_{ i}^{(k+1 \lvert k)}}
\newcommand{\cestatek} {\tilde{\mathbf{s}}_{ i}^{(k)}}

\newcommand{\statekpred} {\hat{\mathbf{s}}_{i}^{(k \lvert k-1)}}
\newcommand{\pstatek} {\mathbf{s}^{(k-1)}}
\newcommand{\fstate} {\mathbf{s}^{(k+1)}}
\newcommand{\Aok} {\mathbf{A}^{(k)}}
\newcommand{\qok} {\mathbf{q}^{(k)}}
\newcommand{\Qok} {\mathbf{Q}^{(k)}}
\newcommand{\Wok} {\mathbf{W}}
\newcommand{\Woxk} {w_{\mathsf{x}}}
\newcommand{\Woyk} {w_{\mathsf{y}}}
\newcommand{\Wozk} {w_{\mathsf{z}}}

\newcommand{\nablapi}{\nabla_{\posi} }
\newcommand{\adirjl} {\mathbf{a}_j^{(\ell_k)}} 
\newcommand{\dik} {d_{i}^{(k)}} 

\newcommand{\djlk}{d_j^{(\ell_k)} }
\newcommand{\phiik}{\phi_i^{(k)}} 
\newcommand{\phijl}{\phi_j^{(\ell_k)}}
\newcommand{\thetaik}{\theta_i^{(k)}} 
\newcommand{\thetajl}{\theta_j^{(\ell_k)}}
\newcommand{\dij} {d_{ij}^{(k)}} 
\newcommand{\dmino} {d_\mathsf{U}^*} 
\newcommand{\dmint} {d_\mathsf{T}^*} 
\newcommand{\vrad}{v_{\mathsf{rad},i}^{(k)}} 
\newcommand{\SNRi}{\mathsf{SNR}_i^{(k)}} 
\newcommand{\SNRo}{\mathsf{SNR}_0}
\newcommand{\fdik}{f_{\mathsf{d},i}^{(k)}} 
\newcommand{\fdjl}{f_{\mathsf{d},j}^{(\ell_k)}} 
\newcommand{\omegax}{\omega_{\mathsf{x}} }
\newcommand{\omegay}{\omega_{\mathsf{y}}}
\newcommand{\omegaz}{\omega_{\mathsf{z}}}
\newcommand{\omegaik}{\bm{\omega}_i^{(k)}} 
\newcommand{\Nmc} {N_\mathsf{MC}}
\newcommand{\Nranging}{\mathcal{N}_\mathsf{r}}
\newcommand{\Nbearing}{\mathcal{N}_\mathsf{b}}
\newcommand{\Njoint}{\mathcal{N}_\mathsf{j}}
\newcommand{\Nall} {\mathcal{N}}
\newcommand{\Nd}{\mathcal{N}_\mathsf{d}}
\newcommand{\Nneighi}{\mathcal{N}_{\mathsf{nb},i}^{(k)}}

\newcommand{\flagrange}{\kappa_j}
\newcommand{\flagbear}{\beta_j}
\newcommand{\flagdop}{\xi_j}
\newcommand{\vzik}{\bm{\mathsf{z}}_i^{(k)}}
\newcommand{\vinfoik}{\bm{\mathsf{i}}_i^{(k)}} 
\newcommand{\zjl}{\mathsf{z}_j^{(\ell_k)}}

\newcommand{\infojl}{\mathsf{i}_j^{(\ell_k)}}
\newcommand{\obsik}{o_i^{(k)}} 

\newcommand{\voik}{\mathbf{w}_i^{(k)}}

\newcommand{\oik}{w_i^{(k)}}
\newcommand{\hik}{h_i^{(k)}}
\newcommand{\nik}{n_i^{(k)}}
\newcommand{\vhik}{\mathbf{h}_i^{(k)}}
\newcommand{\vnik}{\mathbf{n}_i^{(k)}}
\newcommand{\Rik}{\mathbf{R}_i^{(k)}} 
\newcommand{\hikr}{\mathbf{h}_{\mathsf{r},i}^{(k)}}

\newcommand{\hikaz}{\mathbf{h}_{\mathsf{\phi},i}^{(k)}}
\newcommand{\hikel}{\mathbf{h}_{\mathsf{\theta},i}^{(k)}}
\newcommand{\hikdop}{\mathbf{h}_{\mathsf{d},i}^{(k)}}
\newcommand{\stdrangeik}{\sigma_{\mathsf{r},i}^{(k)}}
\newcommand{\sjlr}{\sigma_{\mathsf{r},j}^{(\ell_k)}}

\newcommand{\sikdop}{\sigma_{\mathsf{d},i}^{(k)}}
\newcommand{\sjlaz}{\sigma_{\mathsf{\phi},j}^{(\ell_k)}}
\newcommand{\sjlel}{\sigma_{\mathsf{\theta},j}^{(\ell_k)}}
\newcommand{\sjldop}{\sigma_{\mathsf{d},j}^{(\ell_k)}}
\newcommand{\sdop}{\sigma_{\mathsf{d},0}}
\newcommand{\srange}{\sigma_{\mathsf{r},0}}
\newcommand{\sstb}{\sigma_{\mathsf{b},0}}
\newcommand{\stb}{{\sigma}^2_{\mathsf{b},0}}
\newcommand{\stdik}{\sigma_i^{(k)}}
\newcommand{\stdjl}{\sigma_j^{(\ell_k)}}
\newcommand{\zihistory}{\bm{\mathsf{i}}_i^{(1:k)}}
\newcommand{\infohistory}{\bm{\mathsf{i}}_i^{(1:k)}}
\newcommand{\sjlb}{\sigma_{\mathsf{b},j}^{(\ell_k)}}
\newcommand{\Dx}{  x_{0i}^{(k)}}
\newcommand{\Dy}{  y_{0i}^{(k)}}
\newcommand{\Dz}{  z_{0i}^{(k)}}

\newcommand{\rmax} {r_\mathsf{max}}
\newcommand{\hmax} {h_\mathsf{max}}
\newcommand{\MEANik}{{\mathbf{m}_i^{(k)}}}
\newcommand{\MEANikk}{\mathbf{m}_i^{(k \lvert k)}}
\newcommand{\MEANipred}{\mathbf{m}_i^{(k \lvert k-1)}}

\newcommand{\MEANikpre}{\mathbf{m}_i^{(k-1)}}
\newcommand{\COVikpre}{\mathbf{P}_i^{(k-1)}}

\newcommand{\Sik}{\mathbf{S}_i^{(k)}}
\newcommand{\nablah}{\nabla \mathbf{h}_i^{(k)}}
\newcommand{\nablahp}{\nabla_{\posk} \mathbf{h}_i^{(k)}}
\newcommand{\nablahT}{\nabla^{\mathsf{T}} \mathbf{h}_i^{(k)}}
\newcommand{\nablahpT}{\nabla^{\mathsf{T}}_{\posk} \mathbf{h}_i^{(k)}}
\newcommand{\COVik}{{\mathbf{P}_i^{(k)}}}
\newcommand{\COVikk}{\mathbf{P}_i^{(k \lvert k)}}
\newcommand{\COVipred}{\mathbf{P}_i^{(k \lvert k-1)}}

\newcommand{\Pikpp}{\mathbf{P}_{\mathsf{pp},i}^{(k \lvert k)}}
\newcommand{\Pikpv}{\mathbf{P}_{\mathsf{pv},i}^{(k \lvert k)}}
\newcommand{\Pikvp}{\mathbf{P}_{\mathsf{vp},i}^{(k \lvert k)}}
\newcommand{\Pikvv}{\mathbf{P}_{\mathsf{vv},i}^{(k \lvert k)}}
\newcommand{\FIMik}{\mathbf{J}_i^{(k)}}
\newcommand{\Infikpp}{\mathbf{J}_{\mathsf{pp},i}^{(k \lvert k)}}

\newcommand{\Infikppprev}{\mathbf{J}_{\mathsf{pp},i}^{(k \lvert k-1)}}

\newcommand{\Infikxxprev}{J_{\mathsf{xx},i}^{-}}
\newcommand{\Infikxyprev}{J_{\mathsf{xy},i}^{-}}
\newcommand{\Infikxzprev}{J_{\mathsf{xz},i}^{-}}
\newcommand{\Infikyyprev}{J_{\mathsf{yy},i}^{-}}
\newcommand{\Infikzzprev}{J_{\mathsf{zz},i}^{-}}
\newcommand{\Infikyzprev}{J_{\mathsf{yz},i}^{-}}
\newcommand{\Gjlr}{\mathbf{G}_{\mathsf{r},j}^{(\ell_k)}}
\newcommand{\Gjlaz}{\mathbf{G}_{\mathsf{\phi},j}^{(\ell_k)}}
\newcommand{\Gjlel}{\mathbf{G}_{\mathsf{\theta},j}^{(\ell_k)}}
\newcommand{\Gjldop}{\mathbf{G}_{\mathsf{d},j}^{(\ell_k)}}
\newcommand{\gxx}{g_{\mathsf{xx}}}
\newcommand{\gxy}{g_{\mathsf{xy}}}
\newcommand{\gxz}{g_{\mathsf{xz}}}
\newcommand{\gyy}{g_{\mathsf{yy}}}
\newcommand{\gyz}{g_{\mathsf{yz}}}
\newcommand{\gzz}{g_{\mathsf{zz}}}
\newcommand{\Jxx}{J_{\mathsf{xx},i}}
\newcommand{\Jxy}{J_{\mathsf{xy},i}}
\newcommand{\Jxz}{J_{\mathsf{xz},i}}
\newcommand{\Jyx}{J_{\mathsf{yx},i}}
\newcommand{\Jyy}{J_{\mathsf{yy},i}}
\newcommand{\Jyz}{J_{\mathsf{yz},i}}
\newcommand{\Jzx}{J_{\mathsf{zx},i}}
\newcommand{\Jzy}{J_{\mathsf{zy},i}}
\newcommand{\Jzz}{J_{\mathsf{zz},i}}
\newcommand{\Cxx}{C_{\mathsf{xx}}}
\newcommand{\Cxy}{C_{\mathsf{xy}}}
\newcommand{\Cxz}{C_{\mathsf{xz}}}
\newcommand{\Cyx}{C_{\mathsf{yx}}}
\newcommand{\Cyy}{C_{\mathsf{yy}}}
\newcommand{\Cyz}{C_{\mathsf{yz}}}
\newcommand{\Czx}{C_{\mathsf{zx}}}

\newcommand{\Czz}{C_{\mathsf{zz}}}
\newcommand{\Dik}{\mathcal{D}_i^{(k)}}
\newcommand{\Tpoint} {\Theta_{b}}

\newcommand{\Nchirp} {M_i}

\newcommand{\FMCWfreqsweep}{B_i}
\newcommand{\FMCWtimesweep}{\tau_i}

\newcommand{\Tobs}{T_{i}}
\newcommand{\No}{N_0}
\newcommand{\FMCWPt}{P_{\mathsf{t}}}
\newcommand{\FMCWPn}{P_{\mathsf{n}}}
\newcommand{\kboltz}{\kappa_{\mathsf{b}}}
 \newcommand{\To}{T_0}
 \newcommand{\sph}{\mathsf{s}_{\phi,j}^{(\ell_k)}}
\newcommand{\sth}{\mathsf{s}_{\theta,j}^{(\ell_k)}}
\newcommand{\cph}{\mathsf{c}_{\phi,j}^{(\ell_k)}}
\newcommand{\cth}{\mathsf{c}_{\theta,j}^{(\ell_k)}}

\DeclareMathOperator*{\argmin}{argmin}
\pdfoutput=1
\begin{document}
\bstctlcite{IEEEexample:BSTcontrol}
\bstctlcite{IEEEexample:BSTcontrol}


\title{Dynamic Radar Network of UAVs: \\ A Joint Navigation and Tracking Approach}
\author{Anna Guerra,~\IEEEmembership{Member,~IEEE,} Davide Dardari,~\IEEEmembership{Senior,~IEEE,} \\and Petar M. Djuri\'c,~\IEEEmembership{Fellow,~IEEE,}
\thanks{ 
This project has received funding from the European Union’s Horizon 2020 research and innovation programme under the Marie Sk\l odowska-Curie grant agreement No 793581.
P. M. D. thanks the support of the NSF under Award CCF-1618999. A. G. (corresponding author, anna.guerra3@unibo.it), and D. D.  are with the University of Bologna, Italy. P. M. D. is with ECE, Stony Brook University, Stony Brook, NY 11794, USA.
E-mail: petar.djuric@stonybrook.edu.}
}
\maketitle
\vspace{-0.7cm}
\begin{abstract}
Nowadays there is a growing research interest on the possibility of enriching small flying robots with autonomous sensing and online navigation capabilities. This will enable a large number of applications spanning from remote surveillance to logistics, smarter cities and emergency aid in hazardous environments. 
\noindent In this context, an emerging problem is to track unauthorized small \acp{UAV} hiding behind buildings or concealing in large UAV networks. In contrast with current solutions mainly based on static and on-ground radars, this paper proposes the idea of a dynamic radar network of UAVs for real-time  and high-accuracy tracking of malicious targets. To this end, we describe a solution for real-time navigation of  \acp{UAV}  to  track a dynamic target using heterogeneously sensed information.
Such information is shared by the UAVs with their neighbors via multi-hops, allowing tracking the target by a local Bayesian estimator running at each agent. Since not all the paths are equal in terms of information gathering point-of-view, the UAVs plan their own trajectory by   minimizing the posterior covariance matrix of the target state under \ac{UAV} kinematic and anti-collision constraints. Our results show how a dynamic network of radars attains better localization results compared to a fixed configuration and how the on-board sensor technology impacts the accuracy in tracking a target with different radar cross sections, especially in \ac{NLOS} situations. 
\end{abstract}
\section{Introduction}
\label{sec:intro}
The use of \acp{UAV} in densely inhabited areas like cities is expected to open an unimaginable set of new applications thanks to their low-cost and high flexibility for deployment. They can be useful in response to specific events, like for instance  in natural disasters or terrorist attacks as an emergency network for assisting rescuers \cite{zhao2019uav}, or for extended coverage and capacity of mobile radio networks \cite{shakeri2019design}. In fact, \acp{UAV} have been proposed as flying base stations for future wireless networks \cite{mozaffari2018beyond, gangula2017trajectory, chen2017learning} because 5G and Beyond networks will be characterized by a massive density of nodes requiring high data rates and supporting huge data traffic \cite{cerwall2015ericsson}. This will require a  much higher degree of network flexibility than in the past in order to smoothly and autonomously react to  fast temporal and spatial variations of  traffic demand.
At the same time, the idea of having swarms of \acp{UAV} being accepted by the wide public might be challenging because of the possibility of their  malicious use \cite{guvenc2018detection, koohifar2018autonomous}. 
In fact, an important problem is the possible presence of sinister \acp{UAV} that can hide behind buildings for illegal  activities, e.g., terrorist attacks, or can blind \ac{UAV} swarms to inhibit their functionality. The problem of fast, reliable, and autonomous detection and tracking of malicious \acp{UAV} is challenging and  still an unsolved issue because most solutions would require the deployment of ad-hoc aerial or terrestrial radar or vision-based infrastructures that might not be economically sustainable or acceptable \cite{bisio2018unauthorized}. 
\begin{figure}[t!]
  \psfrag{M}[lc][lc][0.6]{Measurements} 
  \psfrag{R}[lc][lc][0.6]{Radar}
  \psfrag{I}[lc][lc][0.6]{Malicious}
  \psfrag{U}[lc][lc][0.6]{UAV}
  \psfrag{S}[lc][lc][0.6]{Network of}
  \psfrag{UU}[lc][lc][0.6]{UAVs}
  \centering
\includegraphics[width=0.6\textwidth]{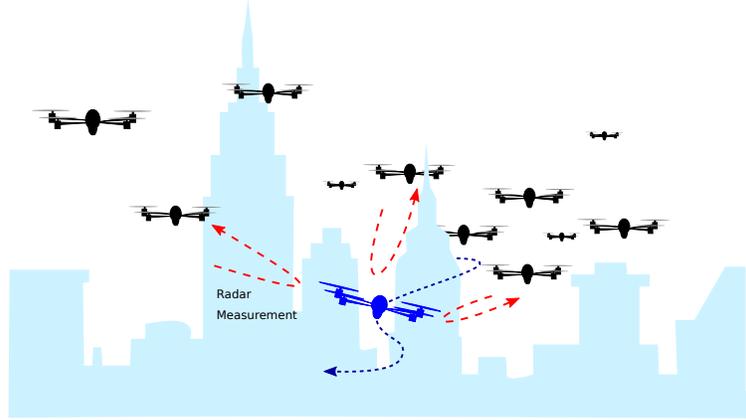}
 \caption{Pictorial representation of a {DRN} considered in this paper.}
\label{fig:radarnetwork}
\end{figure}

Today, current technological solutions  are mainly based on surface-sited (terrestrial) and fixed radars, as battlefield radars, bird detection radars, perimeter surveillance radars, or high-resolution short-range radars, adopted in critical areas (e.g., airports) (see \cite{guvenc2018detection,bisio2018unauthorized, koohifar2018autonomous,hugler2018radar,ezuma2019micro,casbeer2006connectivity} and  the references therein). The possibility of monitoring the movement of small-sized UAVs using a multi-functional airfield radar is considered in \cite{ezuma2019micro}. In \cite{solomitckii2018technologies, paul2015extending,schuster2006performance}, the detection and localization performance of  \ac{FMCW} radar systems is discussed.  
In \cite{casbeer2006connectivity}, a joint connectivity and navigation problem is considered when the radar receiver is mounted on UAVs while the transmitter is on the ground in a multi-static configuration. In \cite{liu2019relative}, a network of \acp{UAV} is used to track ground vehicles.

Nevertheless, the tracking of a malicious UAV with conventional terrestrial radars poses some difficulties since UAVs might be of small size and concealed within the UAV swarm, implying a low probability of being detected and tracked. 
For these reasons, differently from the literature and from our previous works \cite{guerra2018collaborative, guerra2018joint, guerra2019non}, where usually radar sensor networks and \acp{UAV} are treated separately,  this paper aims at introducing the concept of a monostatic \ac{DRN} consisting of \acp{UAV} carrying scanning radars of small sizes and weights, able to track a target and, simultaneously, adapt their formation-navigation control based on the quality of the signals backscattered by a non-cooperative (passive) flying target present in the environment. 
The considered network interrogates the surrounding via echoing signals, estimates and exchanges some target position-related information (e.g., ranging, bearing, and/or Doppler shifts), and jointly infers the target's current position and velocity. The proposed scenario is displayed in Fig.~\ref{fig:radarnetwork} where each \ac{UAV} individually exchanges measurements with neighboring \acp{UAV} and takes navigation decisions on-the-fly in order to reduce the uncertainty on target tracking. 

In order to realize the aforementioned UAV-\ac{DRN}, on-board radar technology should be chosen according to the \ac{UAV} size and maximum payload. To this end, a promising solution might be to use  \ac{mm-wave} radar technology because of the possibility to miniaturize it for an on-board system and for its ranging accuracy and precision thanks to its larger available bandwidth \cite{feger200977,folster2005automotive}. 
Furthermore, a MIMO solution can be employed due to its small size, which will  resul in a highly directional radiation pattern (up to $1$-degree angular accuracy \cite{feger200977}). For example, in \cite{folster2005automotive}, \ac{FMCW} radar sensors working at $77$ GHz are proposed for automotive applications.
Moreover, when considering a target whose size is comparable to that of a mini/micro-UAV, \ac{FMCW} scanning radars are usually preferred compared to pulse radars that perform poorly in localizing small \acp{RCS} \cite{ezuma2019micro}.  For this reason, some research activities have focused on the assessment of the \ac{RCS} values of drones and their impact on the detection performance \cite{guvenc2018detection,hoffmann2016micro}. However, how the target \ac{RCS} affects   tracking accuracy and navigation performance is an open issue. 

Another challenge in the realization of a UAV-\ac{DRN} is the design of optimized paths for the \acp{UAV} to track malicious targets in the best possible way. The optimization of \ac{UAV} trajectories has been the subject of numerous research studies \cite{martinez2006optimal,ragi2013uav,kassas2015receding, dogancay2012uav,wang2019autonomous, cai2019integrated,opromolla2019airborne,ucinski2004optimal,meyer2015distributed,meyer2017scalable,meyer2015distributedob,tang2018autonomous}. In regard to control design, many works in the literature have focused on optimal sensor/anchor placement \cite{martinez2006optimal}, while others tackle the problem from an optimal control point-of-view \cite{tang2018autonomous}.  Among other approaches, information-seeking optimal control (e.g., strategies driven by Shannon or Fisher information measures) has been extensively investigated for localization and tracking applications \cite{ucinski2004optimal,cai2019integrated, dogancay2012uav,meyer2015distributed,meyer2017scalable,meyer2015distributedob,wang2019autonomous}. 
However, these solutions usually do not account for dynamics of the environment and a-priori define the entire paths, and, thus, they are not suitable for our scenario where \acp{UAV} should plan their trajectory in accordance to the movements of the unauthorized flying target. 
\begin{figure}[t!]
\centering
\psfrag{D}[rc][rc][0.6]{\textit{Ranging}}
\psfrag{B}[lc][lc][0.6]{\textit{Bearing}}
\psfrag{J}[lc][lc][0.6]{\textit{Full sensing capabilities}}
\psfrag{d}[c][c][0.5]{$d_i^{(k)}$}
 \psfrag{M}[lc][lc][0.6]{Measurement}
  \psfrag{R}[lc][lc][0.6]{Radar}
 \psfrag{T}[c][c][0.6]{Target}
   \psfrag{x}[c][c][0.6]{$x$}
  \psfrag{y}[c][c][0.6]{$y$}
   \psfrag{z}[c][c][0.6]{$z$}
     \psfrag{p}[lc][lc][0.5]{$\phi_i^{(k)}$}
     \psfrag{t}[lc][lc][0.5]{$\theta_i^{(k)}$}
 \includegraphics[width=0.65\textwidth]{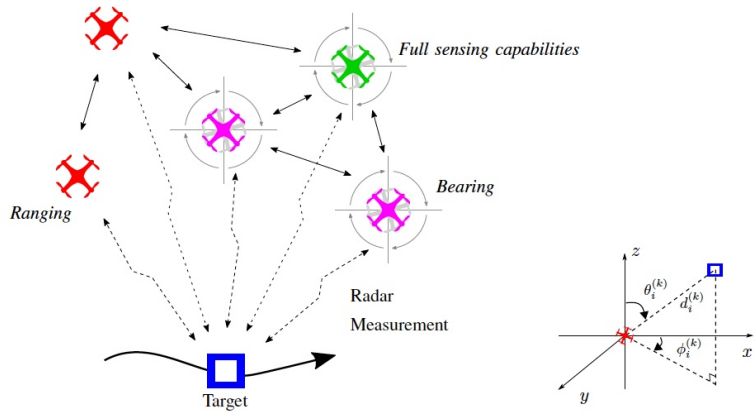}
 \caption{A UAV network, where different groups of  \acp{UAV}  acquire radar measurements. On the left, starting from the environment echo, the red \acp{UAV} estimate ranging information, the magenta the direction of arrival, and the green has full sensing capabilities (including the possibilities of inferring Doppler shifts).  On the right, there is the used coordinate system.} 
\label{fig:network}
\end{figure}

Therefore, the aim of this paper is to study a \ac{UAV} \ac{DRN} as a cooperative radar sensing network for jointly tracking a non-authorized \ac{UAV} in real-time and with  high-accuracy and for smartly navigating the environment in order to reduce the correspondent tracking error 
(via multi-hop exchange of information).  The design of  \acp{DRN}, where the sensors and the target are flying (hence, mobile) poses new challenging issues because of their reconfigurability and mobility, but also offers an unprecedented level of flexibility for target tracking systems thanks to an increased degrees of freedom.  
Since not all the paths are the same from an information gathering point-of-view, the navigation will be formulated as a 3D optimization problem where an information-theoretic cost function permits to combine the a-priori information given by the history of measurements and the contributions brought by the currently acquired  data, that can be delayed by the number of hops (and, hence, they can be aged). 
The impact of the \ac{RCS} of small target (e.g., micro-\ac{UAV}) will be taken into consideration in the measurement noise model, and in assessing the estimation accuracy.

The rest of the paper is organized as follows: Sec. \ref{sec:problemstate} describes the  problem, Sec. \ref{sec:targetlocandtrack} reports details about the radar signal model and the tracking of a non-cooperative UAV, Sec. \ref{sec:costfunction} derives the cost function for optimizing the UAV navigation,  Sec. \ref{sec:controllaw} provides a possible solution for the optimization problem, and Sec. \ref{sec:cs} describes some simulations results.

\textbf{\textit{Notation:}} Vectors and matrices are denoted by bold lowercase and uppercase letters, respectively; $[\mathbf{X}]_{ij}$ denotes the $(i,j)$-th entry of the matrix $\mathbf{X}$; $f\left( x \right)$ symbolizes a probability density function (pdf) of a continuous random variable $x$; $f\left( x \lvert z \right)$ is the conditional distribution of $x$ given $z$; $\mathbf{x} \sim \mathcal{N}\left( \bm{\mu}, \bm{\Sigma} \right)$ means that $\mathbf{x}$ is distributed according to a  Gaussian pdf  with mean $\bm{\mu}$ and covariance matrix $\bm{\Sigma}$; ${x} \sim \mathcal{U}\left[ a, b \right]$ denotes that ${x}$ is a uniform random variable with support $\left[a, b\right]$; $\mathbb{E}\left\{ \cdot \right\}$ represents the expectation of the argument; $\left[ \cdot \right]^{\mathsf{T}}$ denotes  transposition of the argument. Finally, $\mathbf{I}_{n \times m}$ and $\mathbf{0}_{n \times m}$ indicate the identity and zero matrices of $n \times m$ size, respectively.
\section{Problem Statement}
\label{sec:problemstate}
We consider a \ac{DRN} of $N$ \acp{UAV} acting as mobile reference nodes (that is, with  a-priori known positions, for instance available from {GPS}) that navigate through an outdoor environment in order to optimize the accuracy in tracking the position, $
\posk$, and the velocity, $
\velk$, of a moving non-cooperative target. 
The time is discrete and indexed with the symbol $k$. %
\begin{figure*}[t!]
\begin{minipage}{1\textwidth}
\psfrag{M}[lc][lc][0.65]{Radar} 
\psfrag{P}[lc][lc][0.65]{Measurements} 
\psfrag{ED}[lc][lc][0.65]{\textbf{Estimation by diffusion}} 
\psfrag{N}[lc][lc][0.65]{Control Signal} 
\psfrag{S}[lc][lc][0.65]{Estimator, $\mathsf{CE}$} 
\psfrag{C}[lc][lc][0.65]{Combination} 
\psfrag{A}[lc][lc][0.65]{Measurement} 
\psfrag{Ph}[lc][lc][0.65]{Exchange} 
\psfrag{T}[lc][lc][0.65]{Target State} 
\psfrag{L}[lc][lc][0.65]{Estimator, $\mathsf{SE}$} 
\psfrag{TN}[rc][rc][0.65]{\textit{To neighbors}}
\psfrag{D}[rc][rc][0.65]{$\mathsf{i}_{i}^{(k)}$} 
\psfrag{E}[rc][rc][0.65]{$\hat{\mathbf{s}}_{0;i}^{(k)}$} 
\psfrag{DD}[lc][lc][0.65]{$\mathsf{i}_{j}^{(\ell_k)}, j \neq i$} 
\psfrag{EE}[lc][lc][0.65]{$\hat{\mathbf{s}}_{0;j}^{(\ell_k)}, j \neq i$} 
\psfrag{FN}[lc][lc][0.65]{\textit{From neighbors}} 
\psfrag{d}[c][c][0.65]{$\mathsf{z}_{i}^{(k)}$} 
\psfrag{TP}[c][c][0.65]{\quad\quad$\MEANik, \COVik$} 
\psfrag{CT}[lc][lc][0.65]{$ \cestatek, \COVik$} 
\psfrag{NP}[c][c][0.65]{${\mathbf{u}}_{i}^{(k)}$} 
\psfrag{Z}[c][c][0.65]{$\vinfoik=\left[\ldots, \mathsf{i}_j^{\left( \ell_k\right)}, \ldots \right]^{\mathsf{T}}$} 
\psfrag{O}[lc][lc][0.65]{$b_i^{(k)}$} 
\psfrag{UU}[lc][lc][0.65]{$i$-th UAV} 
\centering
\includegraphics[width=0.95\textwidth]{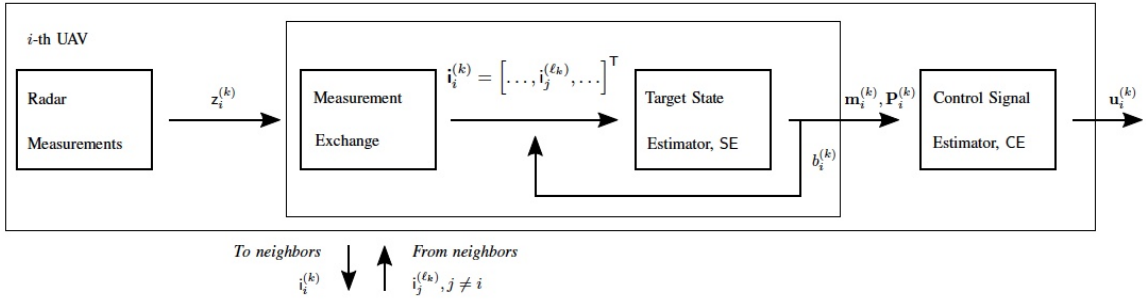}
 \caption{
 A block-diagram for decentralized joint tracking and navigation at the $i$-th \ac{UAV}.}
\label{fig:scheme}
  \end{minipage} 
\end{figure*}

The mobility model of \acp{UAV} can be considered deterministic as the \acp{UAV} are flying outdoors (and, hence, they access the GPS signal with a high degree of accuracy) and, at each time instant, the next {position of the $i$-th \ac{UAV}} is given by
$\fposi=\varphi\left(\posi, \ui \right)$,
where $\varphi\left( \cdot \right)$ is the transition function,  $\posi=\left[\xik, \yik, \zik \right]^{\mathsf{T}}$ is the position of the $i$-th UAV at time instant $k$, and $\ui=\left[\uxi, \uyi, \uzi  \right]^{\mathsf{T}}=g\left( \speedik, \headik, \tiltik \right)$ is the control signal that the $i$-th \ac{UAV} computes on its own for accurate tracking of the target \cite{ragi2013uav}. The magnitude of the speed, the heading and the tilt angles are denoted by  $\speedik$, $\headik$, and $\tiltik$, respectively. In particular, the update of the position is given by
\begin{align}\label{eq:UAVkinematic}
    &\left[\begin{array}{l}
         \fxik   \\
         \fyik  \\
         \fzik
    \end{array}\right] = \left[\begin{array}{l}
         \xik + \uxi   \\
         \yik + \uyi  \\
         \zik + \uzi  
    \end{array}\right] =\left[\begin{array}{l}
         \xik+ {\left(\speedik\cdot \Delta t \right) \, \cos\left( \headik \right) \, \sin\left( \tiltik \right)} \\  
         \yik + {\left(\speedik \cdot \Delta t \right) \, \sin\left( \headik\right) \sin\left( \tiltik \right)}   \\
          \zik + {\left(\speedik \cdot \Delta t \right) \, \cos\left( \tiltik\right) }
    \end{array}\right]\,,
\end{align}
with $\Delta t$ being the time interval between $k$ and $k+1$.
To make the model more realistic, three constraints are added to impose the minimum and maximum speed and a maximum turn rate in both azimuthal and elevation planes \cite{dogancay2012uav}, that are
\begin{equation}\label{eq:k_const}
\begin{cases}
         \,\, \speedmin \le \speedik \le \speedmax,\\
         \,\, \big\lvert {\headik- \Psi_i^{(k-1)}} \big\rvert  \leq \headmax, \\
         \,\, \big\lvert {\tiltik- \Theta_i^{(k-1)}} \big\rvert  \leq \tiltmax, \\
\end{cases}
\end{equation}
where  $v_{\text{min}}$ and $ v_{\text{max}}$ are the minimum and maximum UAV speeds, respectively, and $\headmax$ and $\tiltmax$ are the turn rate limits, respectively. The geometry of the system in depicted in Fig.~\ref{fig:network}.

On the other hand, the target state vector at time instant $k$ is defined as $\statek=\left[ \left( \posk \right)^{\mathsf{T}},\, \left( \velk  \right)^{\mathsf{T}}  \right]^{\mathsf{T}}$,
where the target position expressed in relation to the $i$-th \ac{UAV} position at time instant $k$ is
 \begin{align}
     &\posk \!\!=\!\! \left[ \begin{array}{l}
     \xok \\ \yok\\ \zok \end{array} \right] \!\!=\!\! \left[ \begin{array}{l}
     \!\xik\!+\! \dik\, \sin\left( \thetaik \right) \, \cos\left( \phiik \right)\!\! \\
     \!\yik\!+\!\dik\, \sin\left( \thetaik  \right) \, \sin\left( \phiik \right)\!\! \\
     \!\zik\!+\!\dik\, \cos\left( \thetaik \right)\!\!
     \end{array}
     \right]\!\!,
 \end{align}
where $\dik= \lVert \posk - \posi \rVert_{2}$ is the distance between the $i$-th \ac{UAV} and the target at time instant $k$, and $\velk=\left[\vox, \, \voy, \, \voz \right]^{\mathsf{T}}$ is its velocity.
The state evolves according to the following dynamic model,
\begin{equation}\label{eq:dyn_target}
\fstate= \Aok \, \statek +\qok,
\end{equation}
where $\Aok$ is the transition matrix, which is assumed known, and $\qok \sim \mathcal{N}\left(\mathbf{0},  \Qok \right)$ is the process noise. 

All \acp{UAV} perform radar measurements with respect to the target, and starting from the acquired data, they can estimate Doppler shifts, ranging and/or bearing from which the position and the velocity of the target can finally be estimated at each time step (two-step localization) through cooperation \cite{dardari2015indoor}. In fact, starting from the radar received signals, the Doppler shift and ranging information can be inferred given the  beat frequency estimation  \cite{schuster2006performance}; whereas the \ac{DOA} can be associated with the antenna steering direction. More specifically, \ac{UAV} rotations might be exploited to point the on-board radar antenna in different angular directions and to form a \ac{RSS} pattern after each rotation as in \cite{isaacs2014quadrotor}.  As an alternative, one may consider a MIMO radar system with electronic beamforming capabilities \cite{hugler2018radar}. 
Hence, each \ac{UAV} can process the collected measurements in different ways: we indicate with $\Nranging$ the set of \acp{UAV} acquiring ranging estimates, $\Nd$ the set able to collect Doppler shifts, $\Nbearing$ the set inferring bearing data, and  $\Njoint$ the set able to estimate all the parameters. The network composed of \acp{UAV} with heterogeneous capabilities is indicated with $\Nall = \Nranging \cup \Nd \cup \Nbearing  \cup \Njoint$. 

In accordance with Fig.~\ref{fig:scheme}, the $i$-th \ac{UAV} performs the following steps at time instant $k$:

\paragraph{Measurement step} The first task is to retrieve state-related information from radar measurements, i.e., from the signal backscattered by the environment where the malicious target navigates. In Fig.~\ref{fig:scheme}, we indicate with $\mathsf{z}_i^{(k)}$ the estimates inferred by the $i$-th \ac{UAV} at time instant $k$;
\paragraph{Communication step} Once the $i$-th \ac{UAV} obtains its own estimates, it communicates this information to the neighbors together with its own position $\left(\text{defined as\,\,} \mathsf{i}_i^{(k)}=\left[\mathsf{z}_i^{(k)}, \mathbf{p}_i^{(k)} \right]  \right)$, and it receives back the same data from neighboring \acp{UAV} via multi-hop propagation, i.e., $ \mathsf{i}_j^{(\ell_k)}=\left[\mathsf{z}_j^{(\ell_k)}, \mathbf{p}_j^{(\ell_k)} \right]$, where $\ell_k$ is a time index accounting for the delay due to multi-hops \cite{guerra2018collaborative, guerra2018joint, xu20183d}.  Each node can directly communicate with its neighbors within a radius of length $\rmax$, while for greater distances, the information is delayed by $h_{ij}^{(k)}$ time slots, equal to the number of hops between the $i$-th and $j$-th \ac{UAV} at instant $k$. We indicate with $\Nneighi$ the set of neighbors of the $i$-th \ac{UAV} at time instant $k$.  
Due to multi-hop propagation, the information obtained at each \ac{UAV} can be aged, preventing an updated view of the network. Finally, we gather all the acquired data in $\vinfoik$, which is the vector that contains the estimates and locations of the $i$-th \ac{UAV} and its neighbors.
\paragraph{Target Tracking} Given the measurements and the positions of the other \acp{UAV}, the presence of a malicious target can be detected and its state can be tracked by each \ac{UAV}. A Bayesian estimator can be used to compute the a-posteriori probability distribution of the target state given the acquired information \Big(the belief is denoted with $b_i^{(k)}\left(\statek \right)$ in Fig.~\ref{fig:scheme}\Big). In our case, we adopt an \ac{EKF} algorithm to compute the Gaussian belief of the state as $b_i^{(k)}\left( \statek \right)=f\left( \statek \lvert  \zihistory \right) \overset{\mathrm{EKF}}{=} \mathcal{N} \left(\MEANik, \COVik \right)$, where $\MEANik$ and  $\COVik$ are the conditional mean vector and the covariance matrix of the state and $\zihistory $ is the acquired information by the $i$-th \ac{UAV} up to time instant $k$. The \ac{EKF} filter algorithm produces estimates that minimize the mean-squared estimation error conditioned on the history of acquired information. Consequently, the estimate of the state at time $k$,   $\estatek$, is defined as
the conditional mean $\estatek = \MEANikk= \mathbb{E} \left\{\statek \lvert \zihistory \right\}$.\footnote{The notation in the superscript $^{\left(n \lvert m\right)}$ refers to the estimate at the $n$-th time instant conditioned to information acquired until time instant $m$ \cite{sarkka2013bayesian}.} With reference to Fig.~\ref{fig:scheme}, we can write
\begin{align}\label{eq:statefunction}
    \estatek&=\mathsf{SE}\left( \vinfoik,  b_i^{(k-1)}\left(\mathbf{s}_0^{(k-1)}\right) \right) \overset{\mathrm{EKF}}{=}\mathsf{SE}\left( \vinfoik,  \MEANikpre, \COVikpre \right),
\end{align}
with $\mathsf{SE}\left( \cdot \right)$ being a function describing the state estimator. Subsequently, an approach based on diffusion of information \cite{dedecius2016sequential} can follow the tracking step to further enhance the estimation accuracy.
\paragraph{UAV control step} The last step is the control signal estimation by the $i$-th agent that will allow the \ac{UAV} to reach its next position, $\fposi$, according to a given command, $\ui$. Since the quality of the measurements depends on the \ac{DRN} geometry and target position, the control law should properly change the \ac{UAV} formation and position in order to maximize the quality of the tracking process and, at the same time, take into account physical constraints (e.g., obstacles). 
For this reason, at each time step, each \ac{UAV} searches for the next \ac{UAV} formation that minimizes an information-theoretic cost function at the next time instant, that can be written as \footnote{Here we suppose that the connectivity between nodes is unaltered from time instant $k$ to $k+1$, meaning that the $i$-th \ac{UAV} solves the optimization problem by assuming that, at $k+1$, it will communicate with the same neighbors.}
\begin{align}\label{eq:problem}
\left(\fkposi \right)^{\star}=&\argmin_{\fkposi }  \,\,  \mathcal{C}\left(\FIMik\left(\eeposk,\, \fkposi \right)\right), 
\end{align}
where $\kposi=\left[ \ldots, \posjl, \ldots\right]^\mathsf{T}$, $j \in \Nneighi$, is the vector containing the locations of \acp{UAV} that are neighbors of the  $i$-th \ac{UAV} at time instant $k$ \Big(those belonging to the set $\Nneighi$\Big), $\ell_k=k-h_{ij}^{(k)}+1$ is the time instant associated with the exchanged information due to multi-hops, $\mathcal{C}\left(\cdot \right)$ is a function that will be defined in the sequel, $\FIMik$ is the cost function also defined in the next, and $\eeposk=\left[ \eestatek\right]_{1:3}$ is the predicted target position where $\eestatek$ is derived during the \textit{prediction step} of \eqref{eq:statefunction}. 

Then, recalling the transition model \eqref{eq:UAVkinematic}, the control signal of the $i$-th \ac{UAV} that satisfies \eqref{eq:problem} is given by $\ui= \left[ \left(\fkposi \right)^{\star} \right]_i -\posi$, where $\left[ \cdot \right]_i$ {is an operator that}  picks the $i$-th entry of the optimal formation in \eqref{eq:problem}.

According to the D-optimality criterion described in \cite{ucinski2004optimal}, we choose  the following cost function: 
\begin{equation}\label{eq:costfunction}
\mathcal{C}\left(\FIMik\left( \posk, \kposi \right)\right)=
- \ln \det \left(\FIMik\left( \posk; \kposi  \right)\right),
\end{equation}
where   $\text{det}\left( \cdot \right)$ is the determinant operator, and $\FIMik\left(\posk; \kposi\right)$ is the \textit{information matrix} of the target's location as a function of the current and previous locations of the neighboring \acp{UAV}. 
Following the same principle as in \cite{dogancay2012uav}, we consider the posterior covariance matrix in its inverse (information) form as
\begin{equation}\label{eq:BIMdef}
    \FIMik\left(\posk; \kposi\right) = \left[ \left(\COVikk \right)^{-1} \right]_{11},
\end{equation}
where the operator $\left[ \cdot \right]_{11}$ picks the sub-matrix relative to the target position, and with the covariance matrix defined as
\begin{align}
\COVikk &= 
 \left[ \begin{array}{cc}
 \Pikpp & \Pikpv \\
 \Pikvp& \Pikvv
\end{array}
\right],  
    \end{align}
whose diagonal contains the variances of the position and the velocity estimates. The cost function  defined in \eqref{eq:costfunction} requires knowledge of the actual target position which is the unknown parameter to be estimated, and for this reason \eqref{eq:problem}  is evaluated  at the  position estimate available to the $i$-th \ac{UAV} at time instant $k$.

Finally, we consider that the problem is
subjected to the following set of constraints:
\begin{align}\label{eq:constraints}
\begin{cases}
\,\, \dij \geq \dmino, \,\, \dik \geq \dmint, \,\, \mathcal{T}_i \cap \mathcal{O} =  \varnothing, \\ 
\,\, \speedmin \le \speedik \le \speedmax,\, \,\,\,\,\,\,\,\,\,\,\,\,\quad\,\,\,\,\qquad\quad\,\,\, \\
\lvert \Psi_i^{(k+1)} - \headik \rvert  \leq \headmax,  \\
\lvert  \Theta_i^{(k+1)}-\tiltik \rvert  \leq \tiltmax, 
\end{cases}
\end{align}
for $i,j=1, \ldots, N$, and where $\dij$ is the inter-\ac{UAV} distance, $\dmino$ is the anti-collision safety distance among \acp{UAV}, $\dmint$ {is} the safety distance with respect to the target, {$\mathcal{T}_i$ {is the set of} feasible position points of the trajectory of the $i$-th \ac{UAV}}, and $\mathcal{O}$ {is} the set of obstacles present in the environment from which the \acp{UAV} should keep a safety distance {equal to} $d_{\mathsf{O}}^{*}$. 
\section{UAV-Target Tracking}
\label{sec:targetlocandtrack}
The target tracking aims to estimate the state of the target (e.g., its position and velocity) starting from the received echo signals. In this section, we briefly recall the signal model used by a \ac{FMCW} radar that might be integrated in the \ac{UAV} payload and, then, we focus on a Bayesian filtering method for target tracking. More specifically, we adopt an \ac{EKF} as a tool to solve the tracking problem thanks to its capability of dealing with heterogeneous measurements, statistical characterization of uncertainties, and \ac{UAV} mobility models. 
\subsection{Example of Signal Model for on-board FMCW Radar}
A widely used radar technology for \acp{UAV} is the \ac{FMCW} radar that, differently from pulse radars, interrogates the environment with a signal linearly modulated in frequency (namely, chirp). Sometimes, in order to increase the \ac{SNR} and infer Doppler shift measurements, multiple chirps can be transmitted in a fixed time window (chirp train). Once the signal is received back by the radar, it is combined with a template of the transmitted waveform by a mixer. As a result, different target-related parameters, such as ranging and Doppler shifts, can be inferred by processing the frequency and phase information of the signal at the output of this mixer.
In particular, to retrieve velocity information, it is possible to rely on phase differences between different received chirps, or, directly, on Doppler-shift estimates. If the \ac{FMCW} radar consists of multiple transmitting and receiving antennas (MIMO radar), the angle-of-arrival can be estimated through the measurement of phase differences between the antennas. Another possibility is to exploit the \ac{UAV} rotations: by rotating the on-board antenna towards ad-hoc steering directions, the direction of arrival can be inferred by considering the maximum power of the received echoes. 

A promising solution for \ac{UAV} integration is to operate at millimeter-waves so that  \ac{FMCW} radars can be miniaturized and equipped with multiple antennas. By working at high frequencies, a resolution smaller than a millimeter can be obtained thanks to the higher available bandwidth, up to $4$ GHz at $77$ GHz. 
Example of \ac{FMCW} for \acp{UAV} can be found in \cite{hugler2018radar} and the references therein.

For the following analysis and in order to derive a suitable observation model for the tracking algorithm, it is important to characterize the noise uncertainties of the ranging, bearing, and Doppler shift estimates as inferred by the radar. To this end, the \ac{CRLB} expression, which can be viewed as the minimum variance achievable by an unbiased estimator, can be considered for ranging and Doppler shift estimates, given by \cite{ivashko2013performance,ivashko2015topology} 
\begin{align}\label{eq:CRBrangedoppler}
    &\text{var}\left( \dik \right) \ge \frac{3}{2}\, \left( \frac{2\, c}{\gamma} \right)^2   \frac{1}{\left( 2\, \pi\, \FMCWfreqsweep \right)^2\, \SNRi}, \quad \text{var}\left( \fdik \right) \ge \frac{1}{\left( 2\, \pi \right)^2}  \frac{6}{\Tobs^2\,\, \SNRi},
\end{align}
where $\Tobs=\FMCWtimesweep \, \Nchirp$ is the observation time, $\FMCWtimesweep$ is the time sweep of a single sawtooth, $B_i$ is the frequency sweep, $\Nchirp$ is the number of chirps (processing gain), $\gamma=4$ is the path-loss exponent for two-way (radar) channel, $c$ is the speed-of-light, and the \ac{SNR} is defined as
\begin{equation}
    \SNRi= \frac{\lambda^2\,\FMCWPt\, G^2\left( \Tpoint \right)}{\left( 4 \pi\right)^3 \, \FMCWPn} \times \frac{\rho}{\left( \dik \right)^{\gamma}}  = \SNRo \times \frac{\rho}{\left( \dik \right)^{\gamma}} 
\end{equation}
where $\rho$ is the target \ac{RCS}, $\SNRo$ is the \ac{SNR} evaluated at $d_0=1$ m and $\rho_0=1$ m$^2$, $\lambda$ is the wavelength, $\FMCWPt$  is the transmitted power, $G\left(\Tpoint \right)$ is the antenna gain pointing at $\Tpoint=(\theta_b, \phi_b)$,  $\FMCWPn=\No\, B_i$ is the noise power with $\No = \kboltz \, \To\, F$, $\kboltz$ is the Boltzmann constant, $\To$  is the receiver temperature, and $F$ is the receiver noise figure. 

On the other hand, for the bearing case, we suppose that the  noise uncertainty (in terms of standard deviation) is constant in the azimuthal and elevation planes and coincides with the \ac{HPBW} of the on-board antenna.
\subsection{Observation Model}
\label{sec:obsmodel}
As described in the previous section, starting from the received signal echoes, each \ac{UAV} estimates information about the target state, e.g., the  distance and angle from the target or the Doppler shift. Subsequently, such information is exchanged between \acp{UAV} via multi-hops together with the \ac{UAV} positions. At the end of this communication step, each \ac{UAV} puts together the gathered information, exploitable for target tracking in a vector. 
Let $\vinfoik= \left[ \ldots, \infojl, \ldots \right]^{\mathsf{T}}$ be the information available to the $i$-th \ac{UAV} at time instant $k$, where the generic element $\infojl=\left[\zjl, \mathbf{p}_j^{(\ell_k)} \right]^{\mathsf{T}},$ $j \in \Nneighi$, contains the radar estimates and the position of the $j$-th neighboring \ac{UAV} delayed due to the multi-hop connection with the $i$-th agent. The generic radar measurement can be written as
\begin{equation}\label{eq:measurement}
    \mathsf{z}_i^{(k)}=\pik\, \obsik + \left(1-\pik \right)\, \oik,
\end{equation}
 where $\pik$ is a flag indicating the presence (if any) of a \ac{LOS} link between the $i$-th \ac{UAV} and the target, 
 and $\oik$ is an outlier term due to the presence of multipath components or extremely noisy measurements \cite{petitjean2018pimrc}.
The first term in \eqref{eq:measurement} contains information about the target state, that is
\begin{equation}\label{eq:measurementvector}
    \obsik= \hik \left(\statek \right) + \nik,
\end{equation}
where $\hik$ is a function that relates the data to the target state and whose expression depends on the \ac{UAV} sensing and processing capabilities, i.e.,
 \begin{align}
      & \hik = \begin{cases}
      \frac{\gamma}{2} \dik= \frac{\gamma}{2} \, \Big\lVert \posk -\posi \Big\rVert_2, \,\,\,\,\, \text{if $i \in \Nranging \vee \Njoint$}, \\
      \phiik= \text{tan}^{-1}\left( {\Dy}/{ \Dx} \right),  \quad \text{if $i \in \Nbearing \vee \Njoint$}, \\ 
      \thetaik= \text{cos}^{-1}\left( {\Dz}/{ \dik} \right),\quad \text{if $i \in \Nbearing \vee \Njoint$}, \\
      \fdik=\left({\gamma\, \vrad}/{2\, \lambda} \right),\quad\quad\,\,\,\, \text{if $i \in \Nd \vee \Njoint$},
      \end{cases}
 \end{align}
\noindent where $\dik$, $\phiik$, $\thetaik$, and $\fdik$ are the actual distance, azimuth, elevation, and Doppler shift between the $i$-th \ac{UAV} and the target,
$\vrad$ is the radial velocity,  $\vee$ is the or-operator, and $\Dx=\xok-\xik$, $\Dy=\yok-\yik$, and $\Dz=\zok-\zik$.

The measurement noise in \eqref{eq:measurementvector} is modeled as $\nik \sim \mathcal{N}\left(0, \left( \stdik \right)^2 \right)$, where, in accordance with the type of measurement, the ranging and Doppler shift variances are described by the \ac{CRLB} as in \eqref{eq:CRBrangedoppler}, that can be reformulated as
 \begin{align}\label{eq:rangingerror}
 \left( \stdrangeik \right)^2= \srange^2\, \frac{\left( \dik \right)^{\gamma}}{\rho}, \quad \left( \sikdop \right)^2= \sdop^2\, \frac{\left( \dik \right)^{\gamma}}{\rho},
 \end{align}
where $\srange^2$ and $\sdop^2$ are the variances at the reference distance $\dik=1$ m and with a target \ac{RCS} of $\rho=1$ m$^2$. 
On the contrary, the bearing noise variance is constant with respect to the distance and the target \ac{RCS}, and $\stb$ is related to the radar \ac{HPBW}, as previously stated.

Eq.~\eqref{eq:measurement} can be written in vector form as
\begin{align}\label{eq:obsmod}
&\vzik= \vpik \odot \left( \vhik \left( \statek \right) + \vnik \right) + (1- \vpik) \odot \voik,
\end{align}
where $\odot$ is the Hadamard product, and the noise can be described as $\vnik \sim \mathcal{N}\left(\mathbf{0}, \Rik \right)$ with a covariance matrix given by $\Rik=\text{diag}\left(\ldots, \left( \stdjl  \right)^2, \ldots \right)$.
\subsection{UAV-Target Tracking}
Starting from the transition and measurement model previously described, each \ac{UAV} can perform tracking to estimate the state of the target. 
Within this framework, the main goal of each \ac{UAV} is to infer the full joint posterior probability of the state at time instant $k$, $\statek$, given the available information up to the current time instant, namely $\infohistory$. 

In this context, it is possible to define a probabilistic state-space Markovian model by considering the following statistical models:
\begin{itemize}
    \item \textit{Measurement model}. It describes how the state is related to the available information by the likelihood $f\left( \vinfoik  \lvert \statek \right)=f\left( \vzik  \lvert \statek \right)$, defined by the statistical measurement model in \eqref{eq:obsmod};
    \item \textit{State transition model}. It describes how the state evolves in time, in accordance with the dynamic model in \eqref{eq:dyn_target} and given by  $f\left( \statek \lvert \pstatek \right)$.
\end{itemize}
\begin{figure*}
\begin{align}\label{eq:Gpp}
    \Infikpp \left(\posk; \kposi\right)=&\sum_{j=1}^{\lvert \Nneighi \rvert} \pjl \left[ \frac{\flagrange}{\left(\sjlr \right)^2}\, \Gjlr \left( \posk; \posjl \right) +   \frac{\flagdop}{\left(\sjldop \right)^2}\, \Gjldop \left( \posk; \posjl \right)  \right. \nonumber \\
    &\left.  +\, \flagbear  \left( \frac{1}{\left( \sjlaz \right)^2 }\,  \Gjlaz \left( \posk; \posjl \right) + \frac{1 }{\left( \sjlel \right)^2} \, \Gjlel \left( \posk; \posjl \right)   \right) \, \right],
\end{align}
\hrule
\end{figure*}
Given this state-space model, an \ac{EKF} approach can be used because the observation functions in \eqref{eq:measurement} are non-linear and the noises are Gaussian distributed. 
In this case, each \ac{UAV} performs the two main steps of the \ac{EKF} algorithm: (1) A \textit{prediction step} within which each \ac{UAV} computes the predictive information $\left( \MEANipred, \COVipred \right)$ given a model for the target mobility as in \eqref{eq:dyn_target}; and (2) An  \textit{update step} for updating the mean and covariance $\left( \MEANikk, \COVikk \right)$ once a new measurement becomes available.
The Jacobian matrix $\nablah$ is given by 
\begin{equation}\label{eq:jac}
    \nablah= 
    \left[ \begin{array}{cc}
     \nabla_{\posk} \left(\hikr \right) & \mathbf{0}\\
    \nabla_{\posk} \left(\hikaz \right)& \mathbf{0} \\
    \nabla_{\posk} \left(\hikel \right)& \mathbf{0} \\
    \nabla_{\posk} \left(\hikdop \right)&\nabla_{\velk} \left(\hikdop \right)
    \end{array}
    \right]_{ \statek = \MEANipred},
\end{equation}
where the generic elements in \eqref{eq:jac} are the derivatives of the measurement models in \eqref{eq:measurement} with respect to the state, that is.
\begin{align}\label{eq:derivatives}
&\nabla_{\posk} \frac{\gamma}{2}\,\, \djlk = \frac{\gamma}{2}\, \adirjl \left(\phijl, \thetajl \right), \\
%
&\nabla_{\posk} \phijl =\adirjl \left(\phijl+ \pi/2, \pi/2 \right) / \left( \djlk \, \sin(\thetajl)\right), \\
%
&\nabla_{\posk} \thetajl = \adirjl \left(\phijl, \thetajl+ \pi/2 \right) / \left( \djlk \right), \\
%
&\nabla_{\velk} \fdjl = \frac{\gamma}{2\, \lambda}\, \adirjl\, \left(\phijl, \thetajl \right),
\end{align}
\noindent where $\adirjl =\left[ \cph\, \sth, \sph\, \sth, \cth \right]^{\mathsf{T}}$ is the direction vector and where the following notation has been adopted:  $\mathsf{c}_{\alpha, i}^{(k)}= \cos\left(\alpha_i^{(k)}\right)$, $\mathsf{s}_{\alpha, i}^{(k)}= \sin\left(\alpha_i^{(k)}\right)$ with $\alpha_i^{(k)}$ being the azimuth/elevation angle in the set $\left\{ \phiik, \thetaik \right\}$.
Finally, we have
\begin{align}
    &\nabla_{\xok} \fdjl=  \frac{\gamma}{2\,\lambda} \left(- \sph \, \sth \,  \omegaz + \cth \, \omegay \right)\, , \nonumber\\
    &\nabla_{\yok} \fdjl= \frac{\gamma}{2\,\lambda} \left(\cph \, \sth \,  \omegaz - \cth \, \omegax\right)\, , \nonumber\\
    &\nabla_{\zok} \fdjl=\frac{\gamma}{2\,\lambda} \left(-   \cph \, \sth\,  \omegay + \sph\, \sth\,  \omegax \right) \, ,
\end{align}
where the 3D angular velocity is given by
\begin{align}\label{eq:angvel}
    \omegaik &=\left[\omegax,\, \omegay,\, \omegaz  \right]^{\mathsf{T}} =\frac{\left( \posk - \posi \right) \times \left( \velk - \vvik \right)}{\left( \dik\right)^2},
\end{align}
where $\times$ indicates the cross product between the two vectors.
If a measurement is not available (e.g., when a drone collects only ranging information), the correspondent row is eliminated from \eqref{eq:jac}.
\section{Information-Theoretic Cost Function}
\label{sec:costfunction}
The autonomous control  in \eqref{eq:problem} is designed to estimate the next location of each \ac{UAV} in order to maximize its capability to best track the target, considering also the locations and estimates of the neighboring \acp{UAV}.
The tracking performance mainly depends on the prior information acquired (if present), on the \ac{UAV} network formation (geometry) and on the uncertainty of the collected measurements. 

In this section, we aim at deriving the analytical expression of the \textit{information matrix} $\FIMik\left(  \cdot \right)$ in \eqref{eq:costfunction}.
Starting from the information model described in Sec.~\ref{sec:obsmodel} and from the output of the \ac{EKF}, it is possible to write the {information matrix} for the dynamic scenario as  \cite{martinez2006optimal}
\begin{align}\label{eq:BIM}
     &\FIMik\left(\posk; \kposi\right) =\left[ \mathbf{P}_i^{-} - \mathbf{P}_i^{-}\,\, \nablahT\, \left(\Sik \right)^{-1}  \,  \nablah\, \mathbf{P}_i^{-}  \right]_{11}^{-1},
\end{align}
where $\mathbf{P}_i^{-}=\COVipred$ is the predictive covariance, $\nablahT$ is the Jacobian matrix defined in \eqref{eq:jac}, $\Sik=\nablah\,\mathbf{P}_i^{-}\, \nablahT+\Rik$, and $\Rik$ is the covariance matrix that depends on the statistical characterization of the measurement noise.

Then, according to the matrix inversion lemma \cite{bar2004estimation}, \eqref{eq:BIM} can be reformulated in a more convenient form as
\begin{align}\label{eq:BIM2}
    \FIMik\left(\posk; \kposi\right)\!&= \left[ \left( \mathbf{P}_i^{-} \right)^{-1} +\nablahT \left(\Rik\right)^{-1} \nablah \right]_{11} =\Infikppprev + \Infikpp, 
\end{align}
where $\Infikppprev=\left[\left( \mathbf{P}_i^{-} \right)^{-1} \right]_{11}$ is the sub-block matrix corresponding to the predictive information matrix of the target position, while $\Infikpp$ corresponds to the \ac{FIM} for non-random parameters, that is,
\begin{equation}\label{eq:classicFIM}
    \Infikpp \left(\posk; \kposi\right)=  \nablahpT \left( \Rik\right)^{-1}   \nablahp.
\end{equation}

Equation \eqref{eq:classicFIM} puts in evidence the relation of the information model \Big(encapsulated in $\Rik$\Big) and of the \ac{UAV}-target geometric configuration \Big(in the Jacobian matrix, $\nablahp$\Big) on the localization performance. The deterministic \ac{FIM} depends on the true target position and on the \ac{UAV} locations as known by each \ac{UAV}. Because this information is not available, they are substituted with their estimates. After some computation, it is possible to write \eqref{eq:classicFIM} as in \eqref{eq:Gpp} with $\flagrange=1$ if the $j$-th neighbor of the $i$-th UAV can process ranging information ($j \in \Nranging \vee \Njoint$), otherwise $\flagrange=0$; similarly $\flagbear=1$ if bearing data are available at the $j$-th node ($j \in \Nbearing \vee \Njoint$), and $\flagdop=1$ if $j \in \Nd \vee \Njoint$.
As we can see, \eqref{eq:Gpp} is composed of four main terms, each one carrying the position-related information from the corresponding measurements (ranging/bearing/Doppler). In turn, each term has a geometric component dependent on the \ac{UAV}-target positions (the matrices $\mathbf{G}$) weighted by the measurement uncertainty (the factors $1/\sigma^2$). The latter are the inverse of the diagonal entries in the measurement covariance matrix $\Rik$, and are reported in \eqref{eq:rangingerror}. Thanks to the possibility to discriminate LOS/NLOS situations, we assume that the \acp{UAV} exactly know the values of the coefficients in \eqref{eq:rangingerror}.\footnote{For example, this is possible if an electromagnetic map of the environment is available \cite{dardari2012satellite}.} 
The geometric matrices in \eqref{eq:Gpp} are given by 
\begin{align}\label{eq:rangingG}
&\Gjlr\left(\posk; \posjl\right) = \frac{\gamma^2}{4} \,\nabla_{\posk}^{\mathsf{T}} \left( \djlk \right) \nabla_{\posk} \left( \djlk \right) =\frac{\gamma^2}{4} \,\ { \,\adirjl \left(  \adirjl  \right)^\mathsf{T}}, \\
%
&\Gjlaz\left(\posk; \posjl\right) =\nabla_{\posk}^{\mathsf{T}} \left( \phijl \right) \nabla_{\posk} \left( \phijl \right) =\frac{ \Gjlr \left(\phijl + \pi/2, \pi/2 \right)}{\left(\djlk \, \sin \left(\thetajl \right) \right)^2 },   \\
%
&\Gjlel\left(\posk; \posjl\right)=\nabla_{\posk}^{\mathsf{T}} \left( \thetajl \right) \nabla_{\posk} \left( \thetajl \right) =\frac{\Gjlr \left(\phijl,  \thetajl + \pi/2   \right)}{ \left( \djlk   \right)^2 }, \\
&\Gjldop\left(\posk; \posjl\right)=\frac{\gamma^2}{4\, \lambda^2}\, \nabla_{\posk}^{\mathsf{T}}\left(\fdjl \right)\, \nabla_{\posk}\left(\fdjl \right) \label{eq:Gdhop},
\end{align}
\noindent where the elements of \eqref{eq:Gdhop} are reported in Appendix A. 

When all the \acp{UAV} in $\Nneighi$ are collecting non-informative or ambiguous measurements, for example when all the \acp{UAV} are in NLOS with the target ($\pjl=0$, $\forall j$) or all have a malfunction in their processing capabilities ($\flagrange=\flagbear=\flagdop=0$, $\forall j$), they can rely on the previous state information to compute \eqref{eq:BIM2} and to perform the control task. In fact, in \eqref{eq:BIM2}, when the measurement covariance matrix goes to zero (when $\sigma^2$ in \eqref{eq:rangingerror} $\rightarrow \infty$), the only surviving term is the predictive information matrix $\Infikppprev$. The elements of \eqref{eq:BIM2} are given in Appendix B.

In the next section, a solution for the navigation problem in \eqref{eq:problem} is proposed based on a non-linear programming approach.
\section{Navigation Algorithm}
\label{sec:controllaw}
\begin{figure}[t]
\psfrag{x}[c][c][0.8]{$x$ m}
\psfrag{y}[c][c][0.8]{$y$ m}
\psfrag{z}[c][c][0.8]{$z$ m}
\centerline{\includegraphics[width=0.5\linewidth,draft=false]{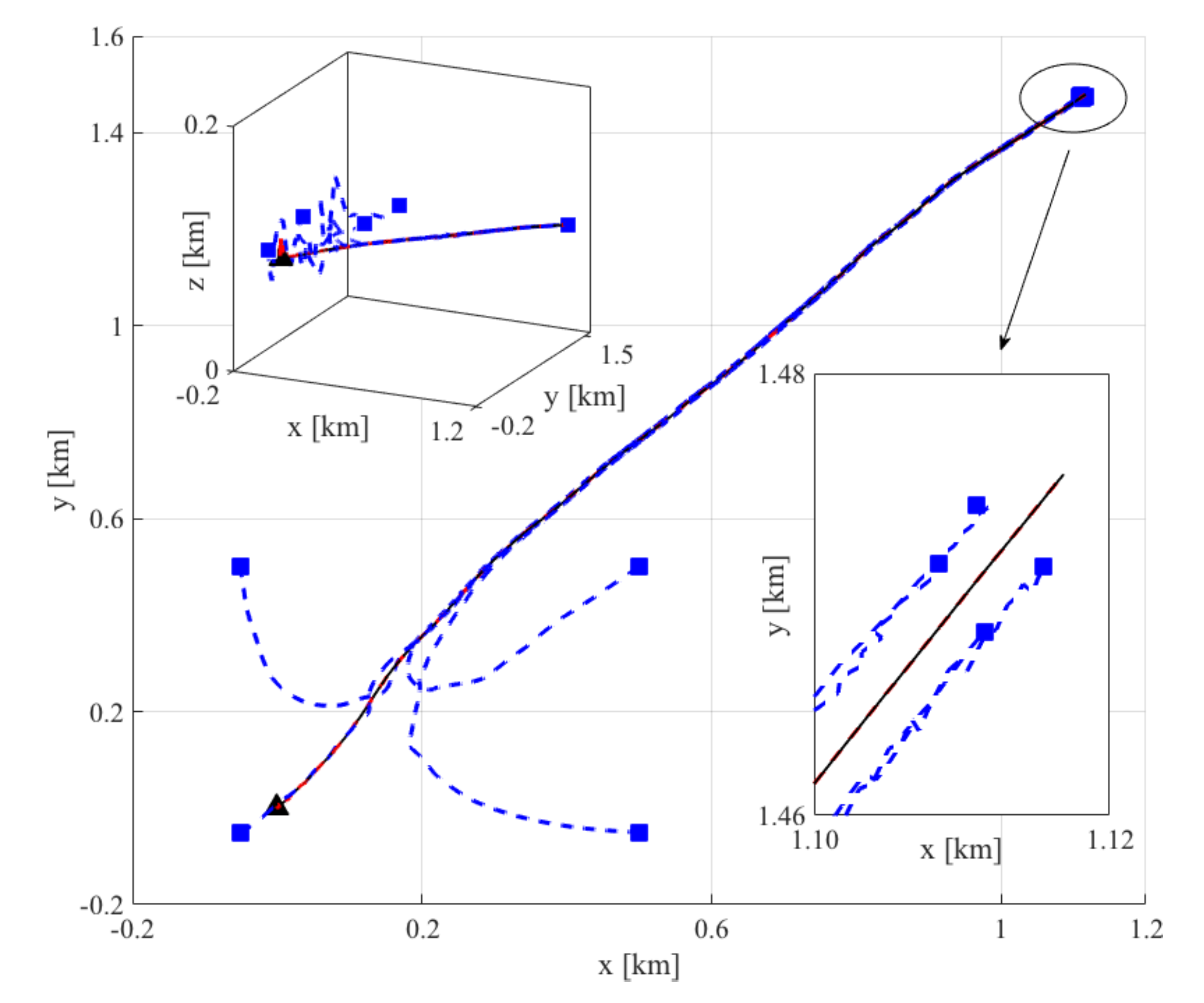} \hspace{0.5cm} \includegraphics[width=0.5\linewidth,draft=false]{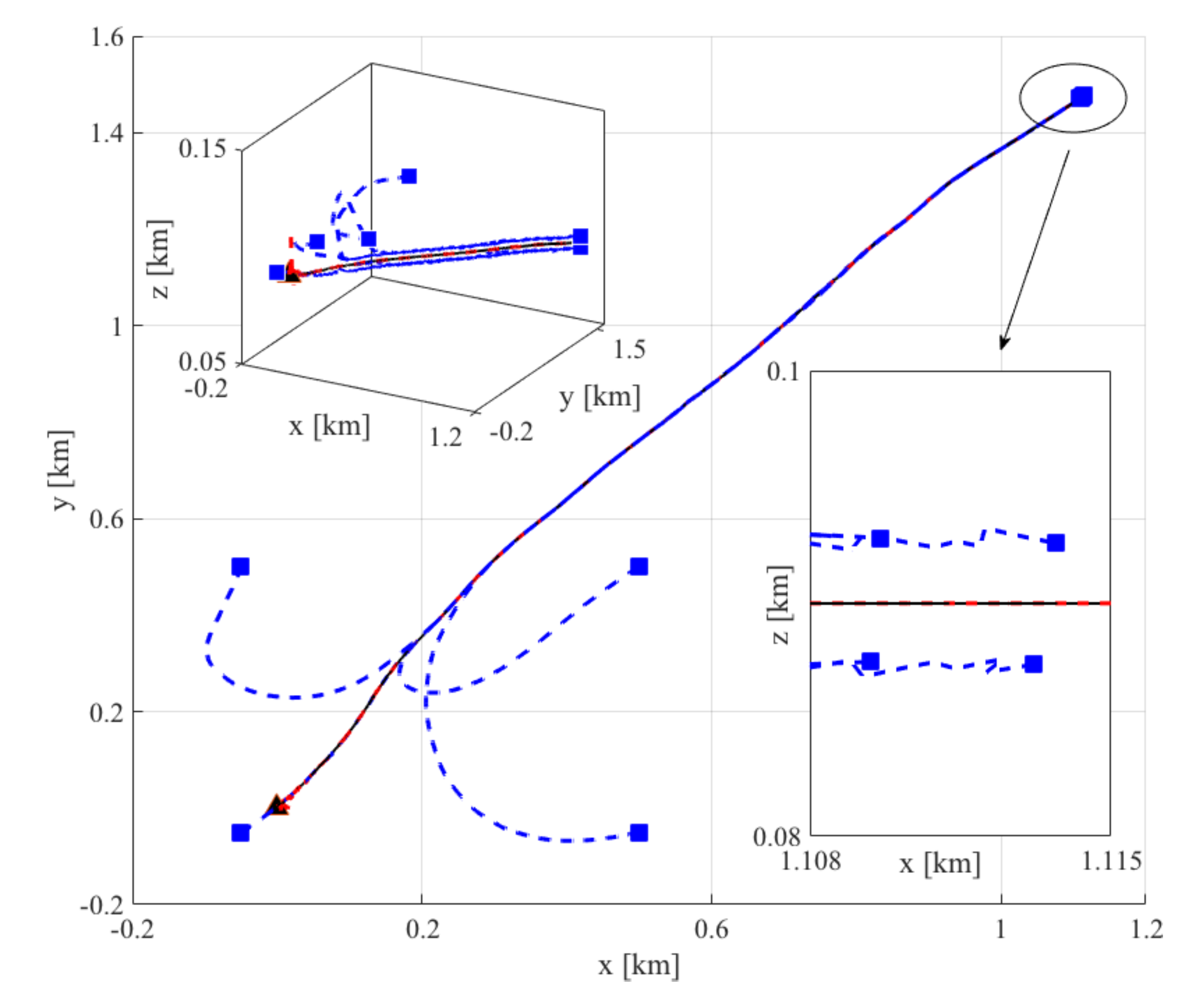}}
\caption{Simulated scenarios in \ac{LOS} conditions with $4$ \acp{UAV} estimating ranging information (left) with an accuracy $\srange=10^{-4}\,$m, and bearing information (right) with an accuracy $\sstb=10^\circ\,$deg. The initial positions and trajectory of the target and \acp{UAV} are indicated with black and blue lines, respectively. The estimated target trajectory is a dashed red line.} 
\label{fig:simulatedscenario_LOS}
\end{figure}
To solve the trajectory problem in \eqref{eq:problem}, one can rely on an approach based on optimization theory (e.g., non-linear programming \cite{luenberger1984linear}) or on a more advanced approaches of machine learning (e.g.,  reinforcement learning algorithms \cite{wang2019autonomous}, dynamic programming \cite{bertsekas1995dynamic} or based on approaches based on graph neural networks \cite{fink2013robust}). 

One possibility to solve the minimization problem in \eqref{eq:problem} is to use a numerical approach as, for example, the projection gradient method  \cite{luenberger1984linear}
\begin{equation}\label{eq:update_g2}
\fui \!=\! -\nu\, \mathbf{P}\, \nabla_{\posi} \, \mathcal{C}\left( \FIMik\left(\eposk,\, \kposi\right)\right) \!-\! \mathbf{N}\left(\mathbf{N}^\mathsf{T} \mathbf{N}\right)^{-1}\mathbf{g},
\end{equation}
where $\nu$ represents the spatial step,  {$\nabla_{\posi} \, \left(\cdot \right)$} is the gradient operator with respect to the \ac{UAV} positions which, taken with the negative sign, represents the direction of decrease of the cost function. The control signal computations are reported in Appendix C.
The projection matrix is denoted with $\mathbf{P}=\mathbf{I}-\mathbf{N}\left(\mathbf{N}^\mathsf{T} \mathbf{N}\right)^{-1}\mathbf{N}^\mathsf{T}$ with $\mathbf{I}$ being the identity matrix and $\mathbf{N}=\left( \nabla_{\posi}\mathbf{g} \right)$ being the gradient of the constraints in $\mathbf{g}=\left[\mathbf{g}_{1} \,\, \mathbf{g}_{2} \,\, \mathbf{g}_{3} \right]$, where 
\begin{align}
&\mathbf{g}_{1}=\mathbf{d}_{\mathsf{U}}-\dmino, \quad \mathbf{d}_{\mathsf{U}}=\left\{\dij: \dij< \dmino\right\}, \\
&\mathbf{g}_{2}=\mathbf{d}_{\mathsf{S}}-\dmint, \,\,\,\,\,\, \mathbf{d}_{\mathsf{S}}=\left\{\dik: \dik< \dmint\right\},  \\
&\mathbf{g}_{3}=\mathbf{d}_{\mathsf{O}}-d_{\mathsf{O}}^*, \,\,\,\,\, \mathbf{d}_{\mathsf{O}}=\left\{d_{i, \mathsf{O}}^{(k)}: d_{i, \mathsf{O}}^{(k)}< d_{\mathsf{O}}^*\right\}.
\end{align}
Finally, we limit the \ac{UAV} speed, altitude and the maximum turning rates according to \eqref{eq:constraints}.
\begin{figure}[t]
\psfrag{NNNNN4Ranging1mm}[lc][lc][0.65]{$N=4$, $\srange=10^{-3}$ m}
\psfrag{N6R1mm}[lc][lc][0.65]{$N=6$, $\srange=10^{-3}$ m}
\psfrag{N10R1mm}[lc][lc][0.65]{$N=10$, $\srange=10^{-3}$ m}
\psfrag{N4R1dm}[lc][lc][0.65]{$N=4$, $\srange=10^{-2}$ m}
\psfrag{N6R1dm}[lc][lc][0.65]{$N=6$, $\srange=10^{-2}$ m}
\psfrag{N10R1dm}[lc][lc][0.65]{$N=10$, $\srange=10^{-2}$ m}
\psfrag{NNNNNNN4Bearing5}[lc][lc][0.65]{$N=4$, $\sstb=5$ deg}
\psfrag{N6Bearing5}[lc][lc][0.65]{$N=6$, $\sstb=5$ deg}
\psfrag{N10Bearing5}[lc][lc][0.65]{$N=10$, $\sstb=5$ deg}
\psfrag{N4Bearing20}[lc][lc][0.65]{$N=4$, $\sstb=20$ deg}
\psfrag{N6Bearing20}[lc][lc][0.65]{$N=6$, $\sstb=20$ deg}
\psfrag{N10Bearing20}[lc][lc][0.65]{$N=10$, $\sstb=20$ deg}
\psfrag{SuccessRate}[c][c][0.8]{Success Rate, $\mathsf{SR}\left( e_{\mathsf{th}} \right)$}
\psfrag{PositioningError}[c][c][0.8]{Positioning Error, $e_{\mathsf{th}}$, [m]}
\psfrag{RangingR4N10aaaaaaaaaa}[lc][lc][0.7]{$N=10$, $\srange=10^{-4}$ m}
\psfrag{RangingR2N10}[lc][lc][0.65]{$N=10$, $\srange=10^{-2}$ m}
\psfrag{RangingR4N6}[lc][lc][0.65]{$N=6$, $\srange=10^{-4}$ m}
\psfrag{RangingR2N6}[lc][lc][0.65]{$N=6$, $\srange=10^{-2}$ m}
\psfrag{RangingR4N4}[lc][lc][0.65]{$N=4$, $\srange=10^{-4}$ m}
\psfrag{RangingR2N4}[lc][lc][0.7]{$N=4$, $\srange=10^{-2}$ m}
\psfrag{BearingB5N10aaaaaaaaaaa}[lc][lc][0.65]{$N=10$, $\srange=5$ deg.}
\psfrag{BearingB20N10}[lc][lc][0.65]{$N=10$, $\sstb=20$ deg.}
\psfrag{BearingB5N6}[lc][lc][0.65]{$N=6$, $\sstb=5$ deg.}
\psfrag{BearingB20N6}[lc][lc][0.65]{$N=6$, $\sstb=20$ deg.}
\psfrag{BearingB5N4}[lc][lc][0.65]{$N=4$, $\sstb=5$ deg.}
\psfrag{BearingB20N4}[lc][lc][0.65]{$N=4$, $\sstb=20$ deg.}
\centerline{\includegraphics[width=0.5\linewidth,draft=false]{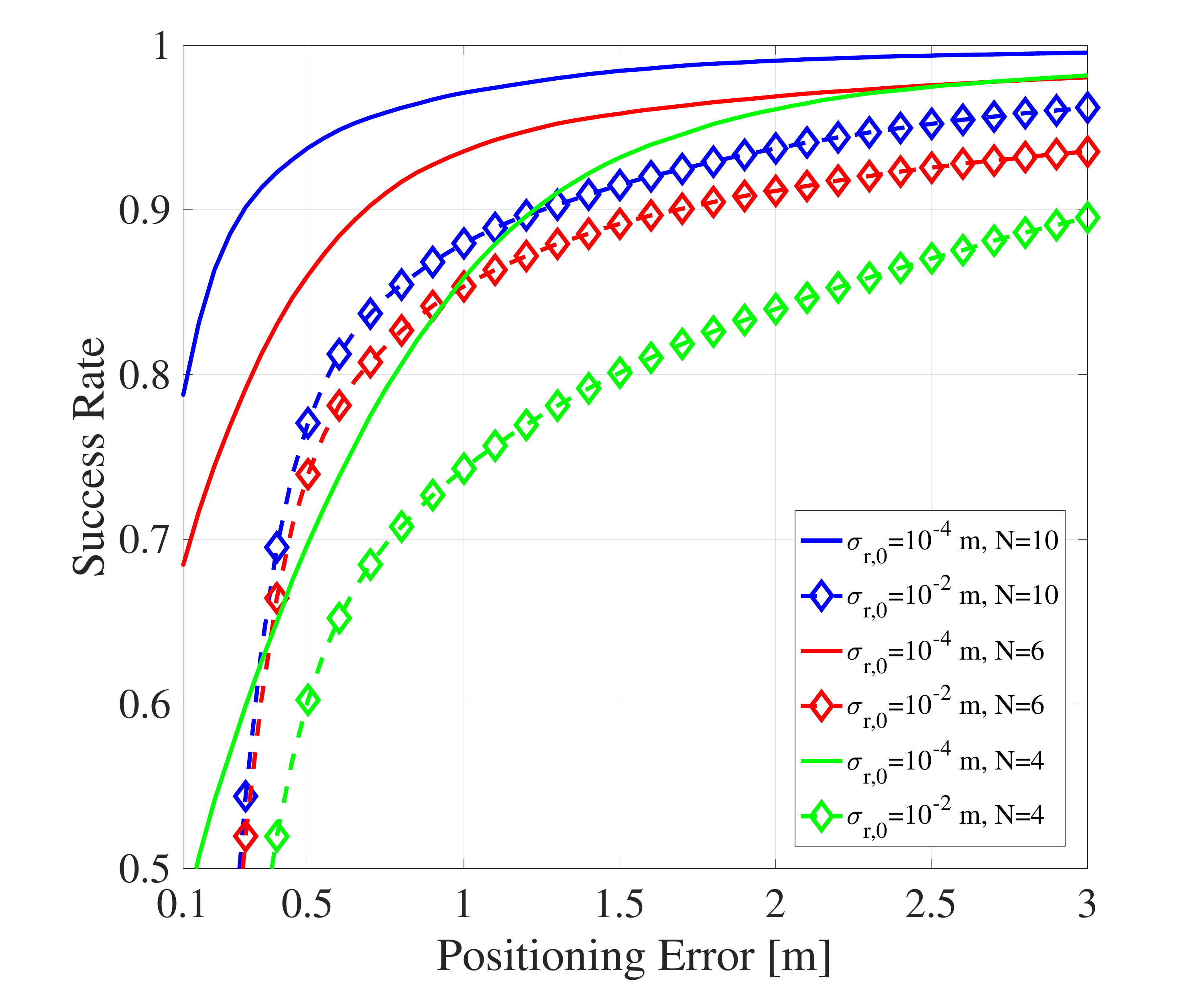} \includegraphics[width=0.5\linewidth,draft=false]{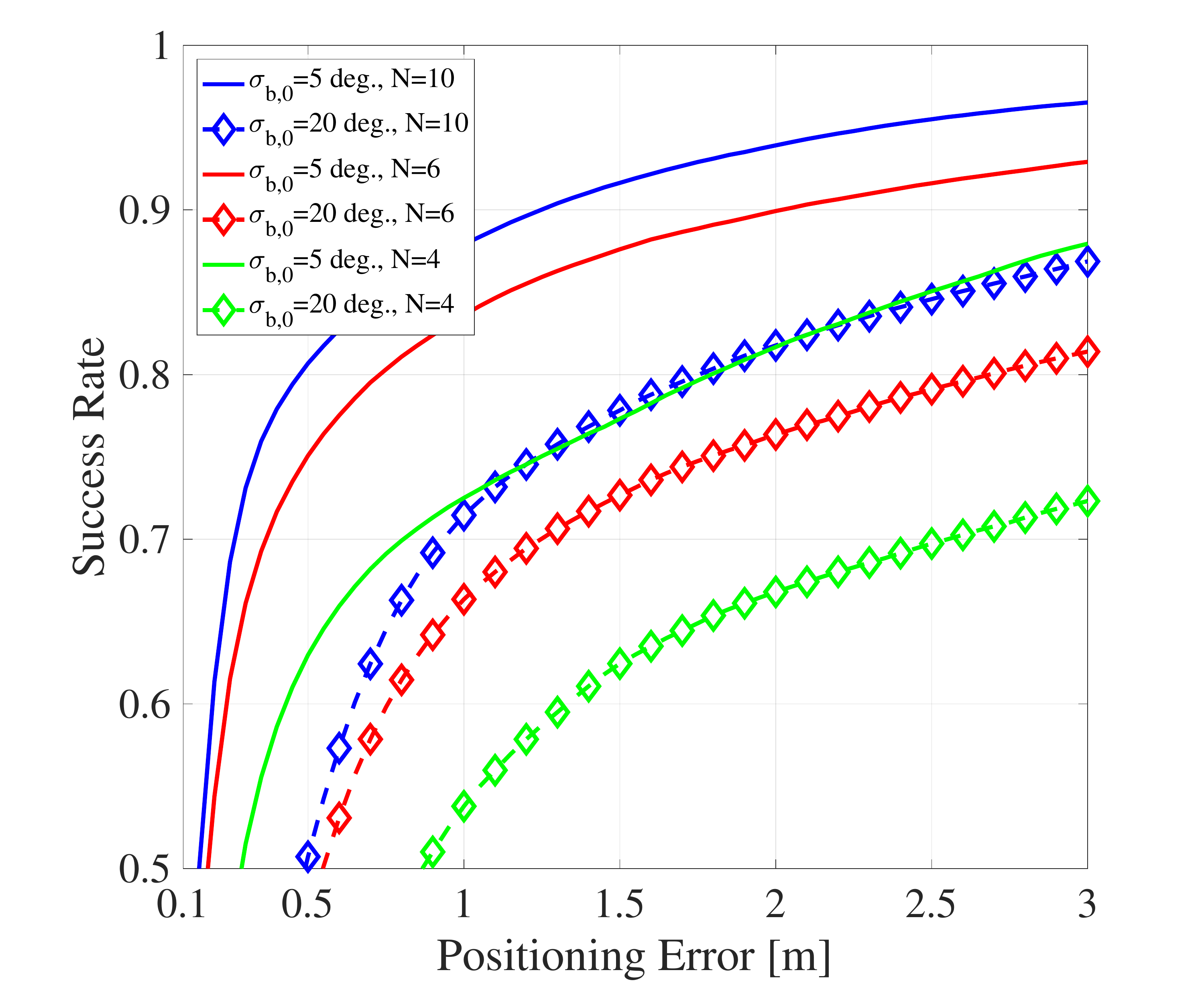}}
\caption{Success rate as a function of the number of UAVs for ranging (left) and bearing (right) measurements and different sensing accuracy.} 
\label{fig:results_SR_N}
\end{figure}

\section{Case Study}
\label{sec:cs}
\begin{figure}[t]
\psfrag{SuccessRate}[c][c][0.8]{Success Rate, $\mathsf{SR}\left( e_{\mathsf{th}} \right)$}
\psfrag{PositioningError}[c][c][0.8]{Positioning Error, $e_{\mathsf{th}}$, [m]}
\psfrag{RangingR5N6aa}[lc][lc][0.65]{$\srange=10^{-5}$ m}
\psfrag{RangingR4N6}[lc][lc][0.65]{$\srange=10^{-4}$ m}
\psfrag{RangingR3N6}[lc][lc][0.65]{$\srange=10^{-3}$ m}
\psfrag{RangingR2N6}[lc][lc][0.65]{$\srange=10^{-2}$ m}
\psfrag{RangingR1N6}[lc][lc][0.65]{$\srange=10^{-1}$ m}
\psfrag{BearingB1N6}[lc][lc][0.65]{$\sstb=1$ deg.}
\psfrag{BearingB5N6}[lc][lc][0.65]{$\sstb=5$ deg.}
\psfrag{BearingB10N6}[lc][lc][0.65]{$\sstb=10$ deg.}
\psfrag{BearingB15N6}[lc][lc][0.65]{$\sstb=15$ deg.}
\psfrag{BearingB20N6}[lc][lc][0.65]{$\sstb=20$ deg.}
 \centerline{\includegraphics[width=0.5\linewidth,draft=false]{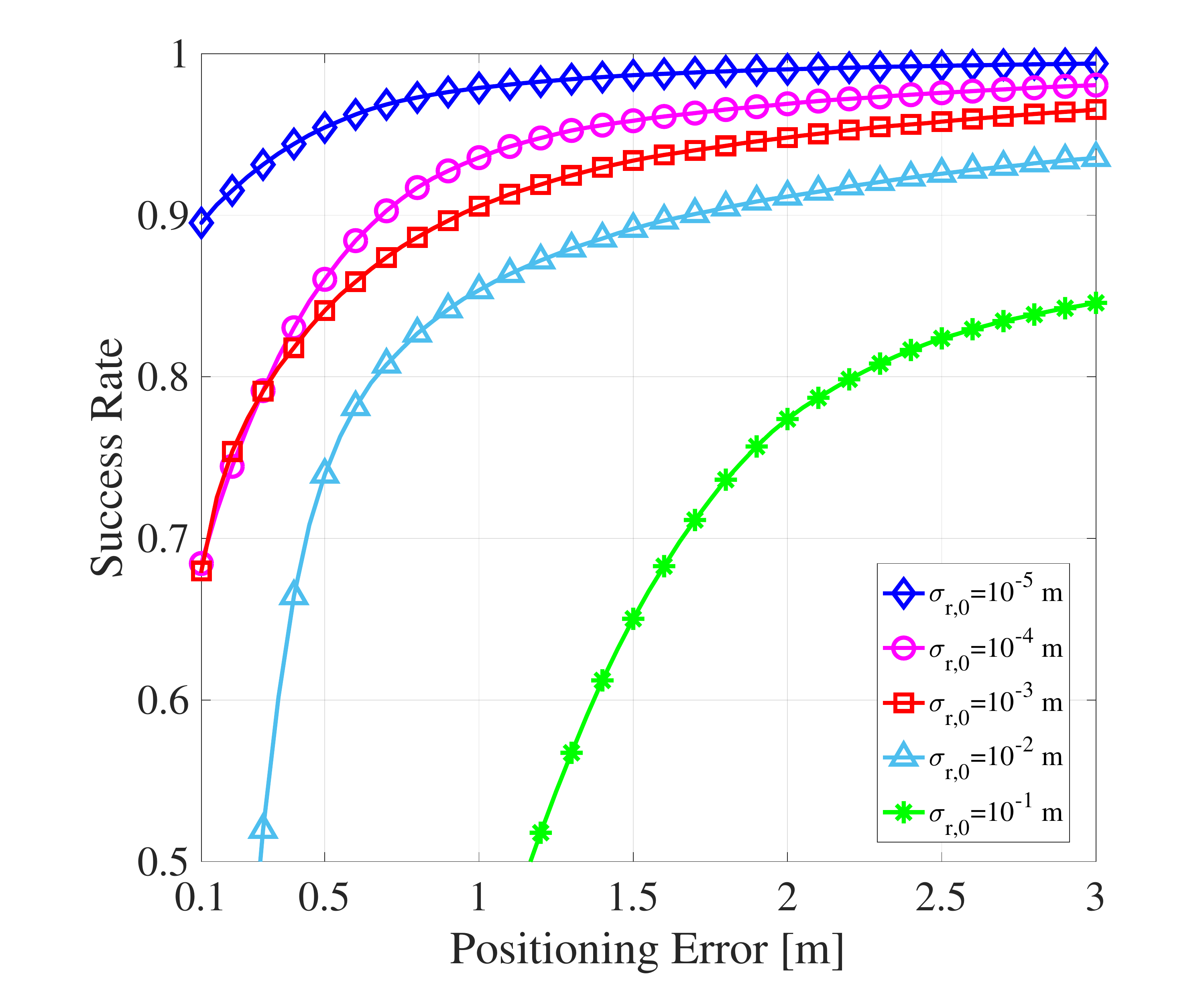}\includegraphics[width=0.5\linewidth,draft=false]{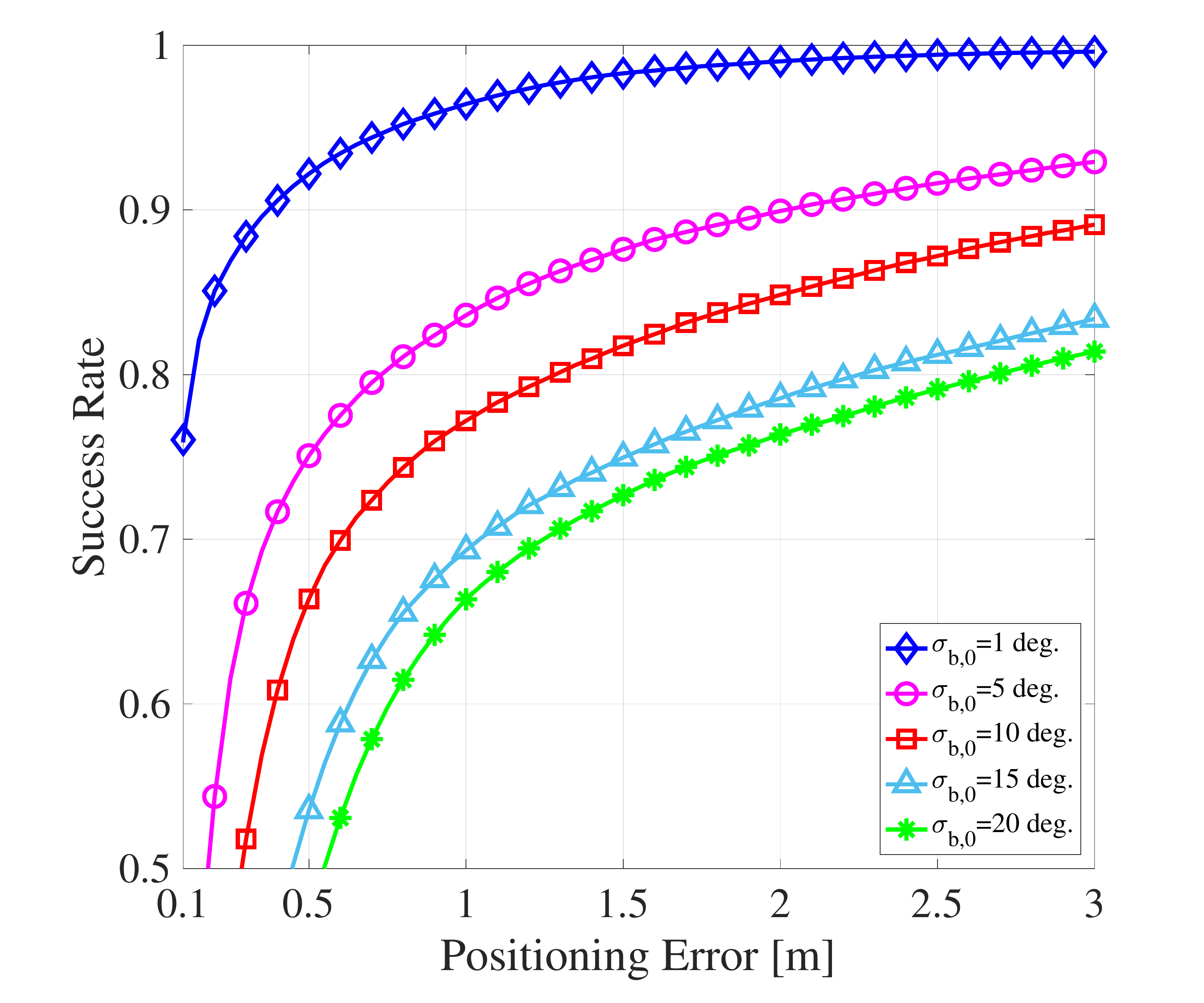}}
\caption{Success rate as a function of the ranging and bearing errors. } 
\label{fig:results_SR}
\end{figure}

\begin{figure}[t]
\psfrag{RangingR1N6}[lc][lc][0.65]{Ranging-only, $\srange=10^{-1}$ m}
\psfrag{RangingR3N6}[lc][lc][0.65]{Ranging-only, $\srange=10^{-3}$ m}
\psfrag{RangingR1N6M64}[lc][lc][0.65]{With Doppler,  $\Nchirp=64$, $\srange=10^{-1}$ m}
\psfrag{RangingR3N6M64}[lc][lc][0.65]{With Doppler,  $\Nchirp=64$, $\srange=10^{-3}$ m}
\psfrag{RangingR1N6M256aaaaaaaaaaaaaaaaaa}[lc][lc][0.65]{With Doppler,  $\Nchirp=256$, $\srange=10^{-1}$ m}
\psfrag{RangingR3N6M256}[lc][lc][0.65]{With Doppler,  $\Nchirp=256$, $\srange=10^{-3}$ m}
\psfrag{RangingR1N6RCS1}[lc][lc][0.65]{$\rho=10^{0}$ m$^2$, $\srange=10^{-1}$ m}
\psfrag{RangingR1N6RCS01}[lc][lc][0.65]{$\rho=10^{-1}$ m$^2$, $\srange=10^{-1}$ m}
\psfrag{RangingR1N6RCS001}[lc][lc][0.65]{$\rho=10^{-2}$ m$^2$, $\srange=10^{-1}$ m}
\psfrag{RangingR1N6RCS0001aaaaa}[lc][lc][0.65]{$\rho=10^{-3}$ m$^2$, $\srange=10^{-1}$ m}
\psfrag{RangingR3N6RCS1}[lc][lc][0.65]{$\rho=10^{0}$ m$^2$, $\srange=10^{-3}$ m}
\psfrag{RangingR3N6RCS01}[lc][lc][0.65]{$\rho=10^{-1}$ m$^2$, $\srange=10^{-3}$ m}
\psfrag{RangingR3N6RCS001}[lc][lc][0.65]{$\rho=10^{-2}$ m$^2$, $\srange=10^{-3}$ m}
\psfrag{RangingR3N6RCS0001}[lc][lc][0.65]{$\rho=10^{-3}$ m$^2$, $\srange=10^{-3}$ m}
\psfrag{SuccessRate}[c][c][0.8]{Success Rate, $\mathsf{SR}\left( e_{\mathsf{th}} \right)$}
\psfrag{PositioningError}[c][c][0.8]{Positioning Error, $e_{\mathsf{th}}$, [m]}
\centerline{\includegraphics[width=0.5\linewidth,draft=false]{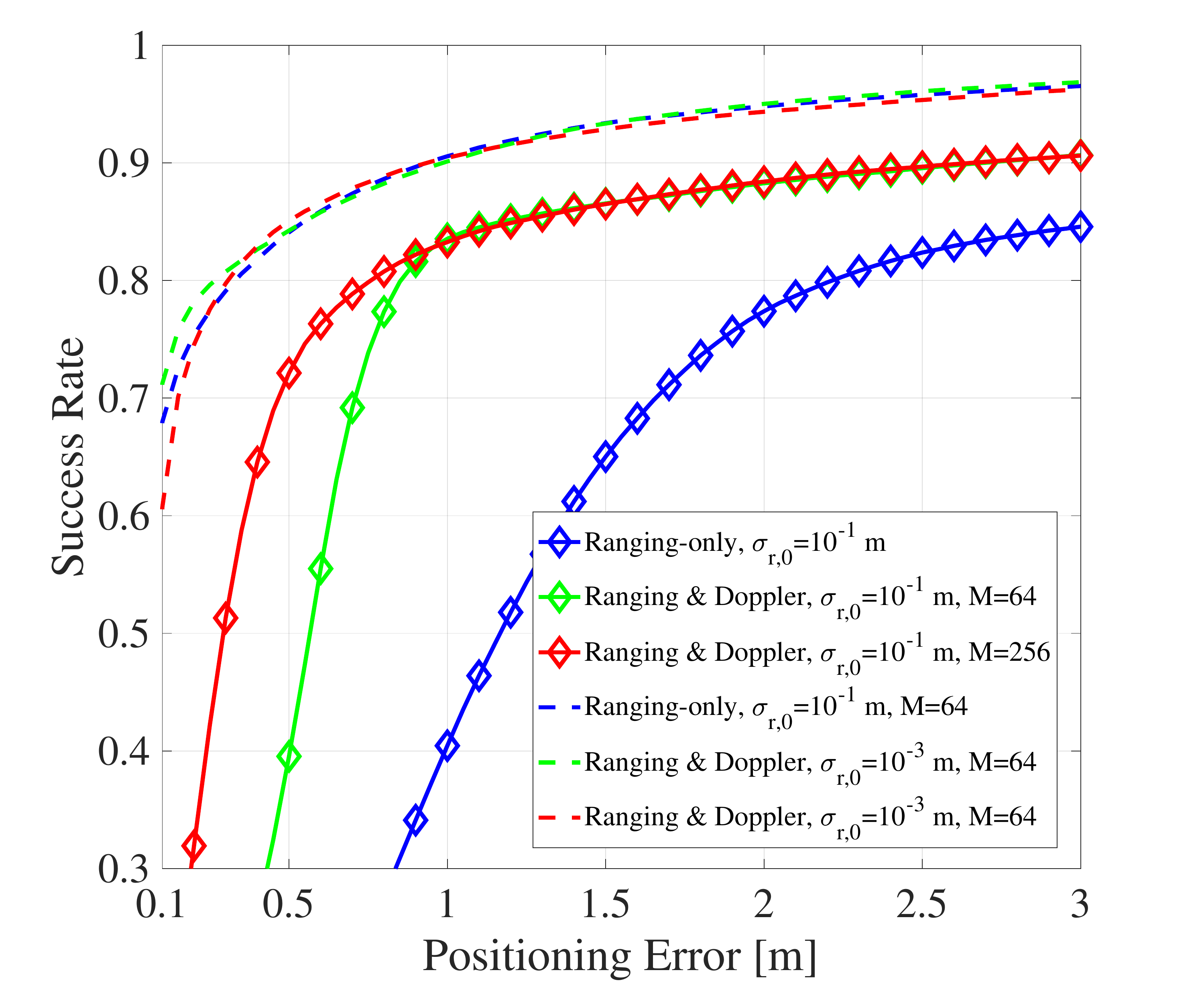}\includegraphics[width=0.5\linewidth,draft=false]{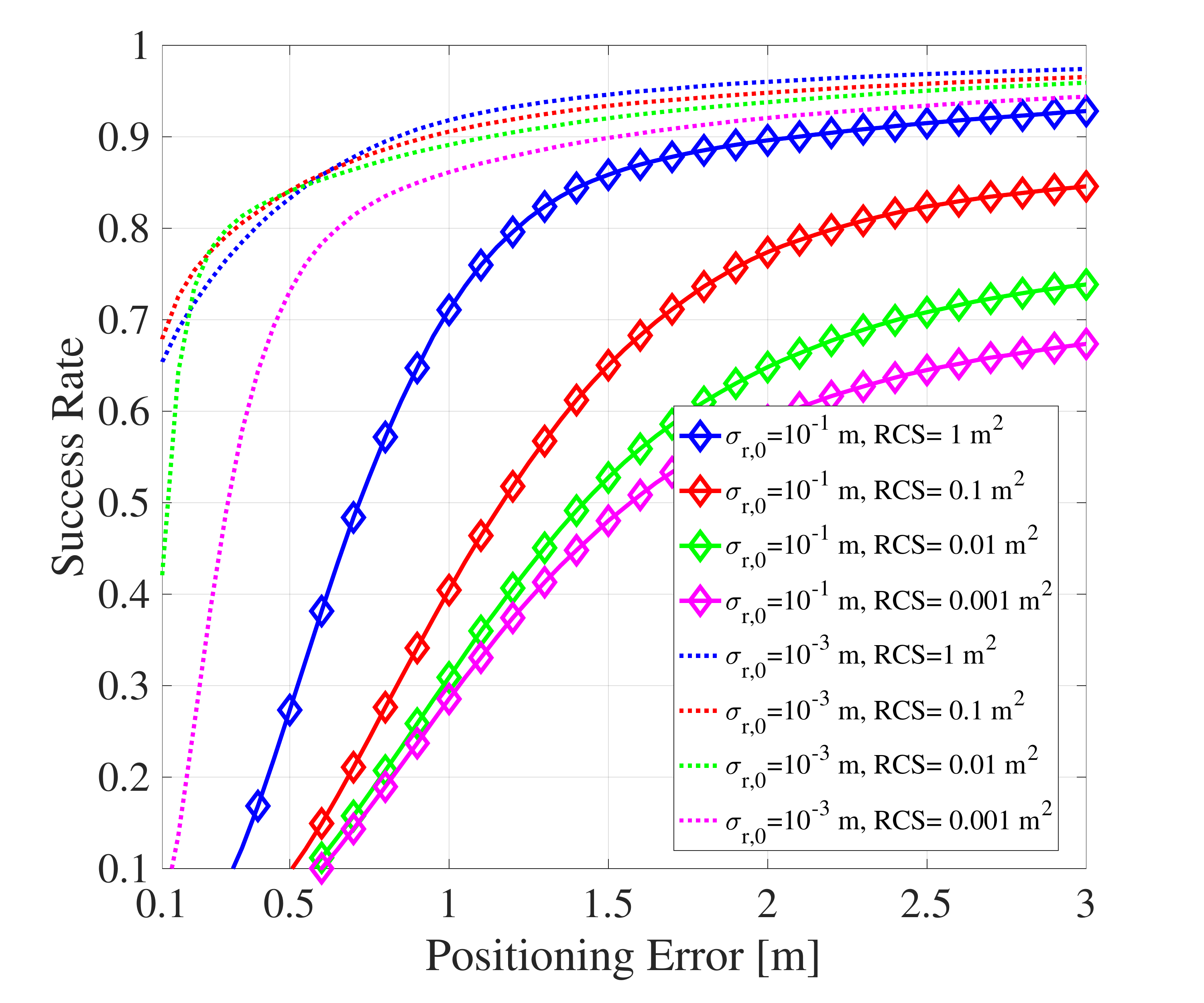}}
\caption{Success rate as a function of the positioning error when ranging measurements are collected. On the left, the case with and without Doppler shifts is considered for a ranging error of $\srange=10^{-3}\,$m (dashed lines) and $\srange=10^{-1}\,$m (diamond markers); on the right, the performance is compared as a function of the RCS.} 
\label{fig:results_SR_RCS}
\end{figure}

\begin{figure}[t]
\psfrag{UAV1}[lc][lc][0.5]{UAV 1}
\psfrag{UAV2}[rc][rc][0.5]{UAV 2}
\psfrag{UAV3}[c][c][0.5]{UAV 3}
\psfrag{UAV4}[lc][lc][0.5]{UAV 4}
\psfrag{SuccessRate}[c][c][0.8]{Success Rate, $\mathsf{SR}\left( e_{\mathsf{th}} \right)$}
\psfrag{PositioningError}[c][c][0.8]{Positioning Error, $e_{\mathsf{th}}$, [m]}
\psfrag{centralized}[lc][lc][0.5]{$\rmax=\infty$}
\psfrag{Hmax3aaaaaaaaaaaaaa}[lc][lc][0.5]{$\hmax=3$, $\rmax=505$ m}
\psfrag{Hmax2}[lc][lc][0.5]{$\hmax=2$, $\rmax=505$ m}
\psfrag{Hmax1}[lc][lc][0.5]{$\hmax=1$, $\rmax=505$ m}
\psfrag{x}[lc][lc][0.65]{$x$ [m]}
\psfrag{y}[lc][lc][0.65]{$y$ [m]}
\centerline{
\includegraphics[width=0.3\linewidth,draft=false]{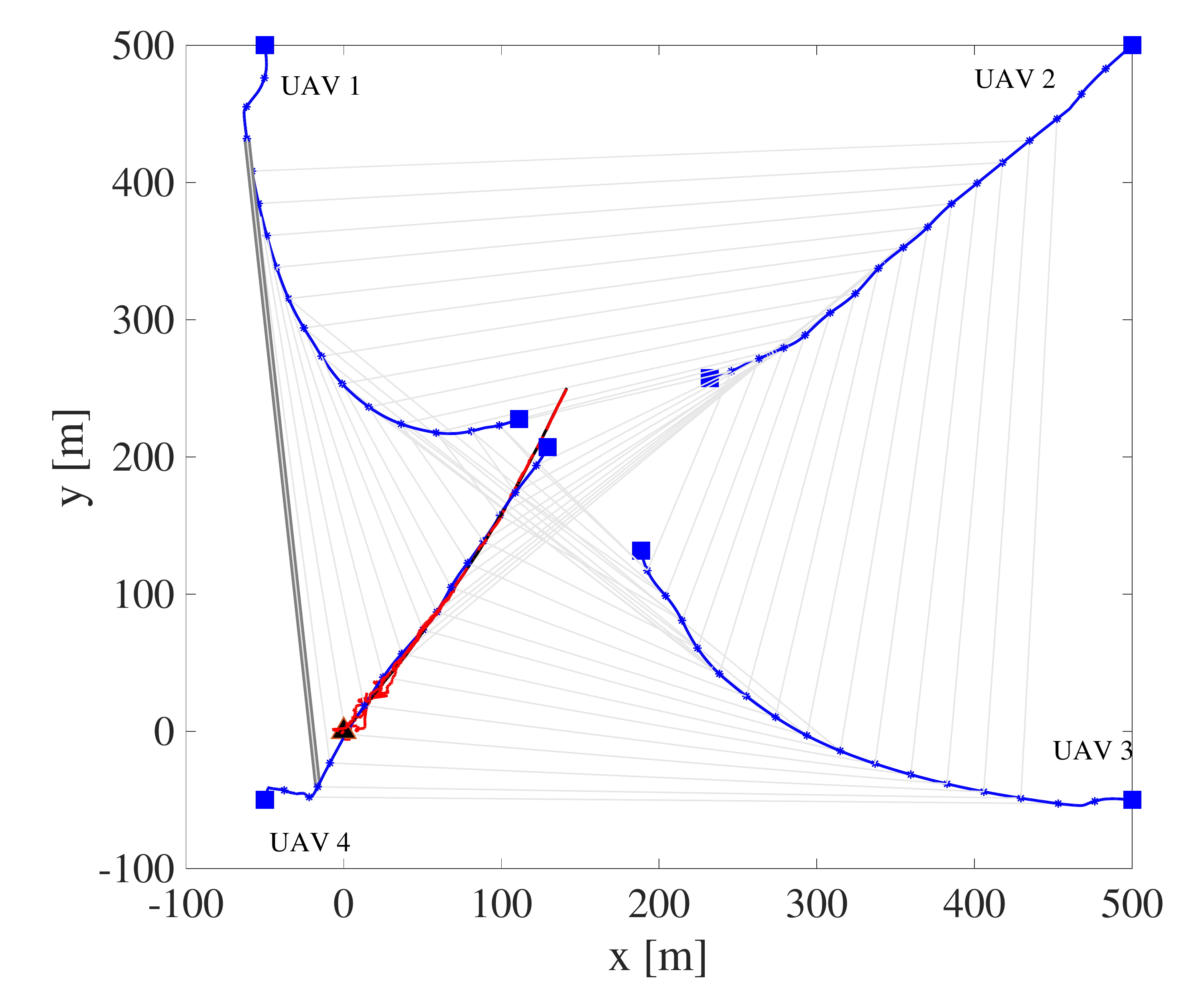}
\includegraphics[width=0.3\linewidth,draft=false]{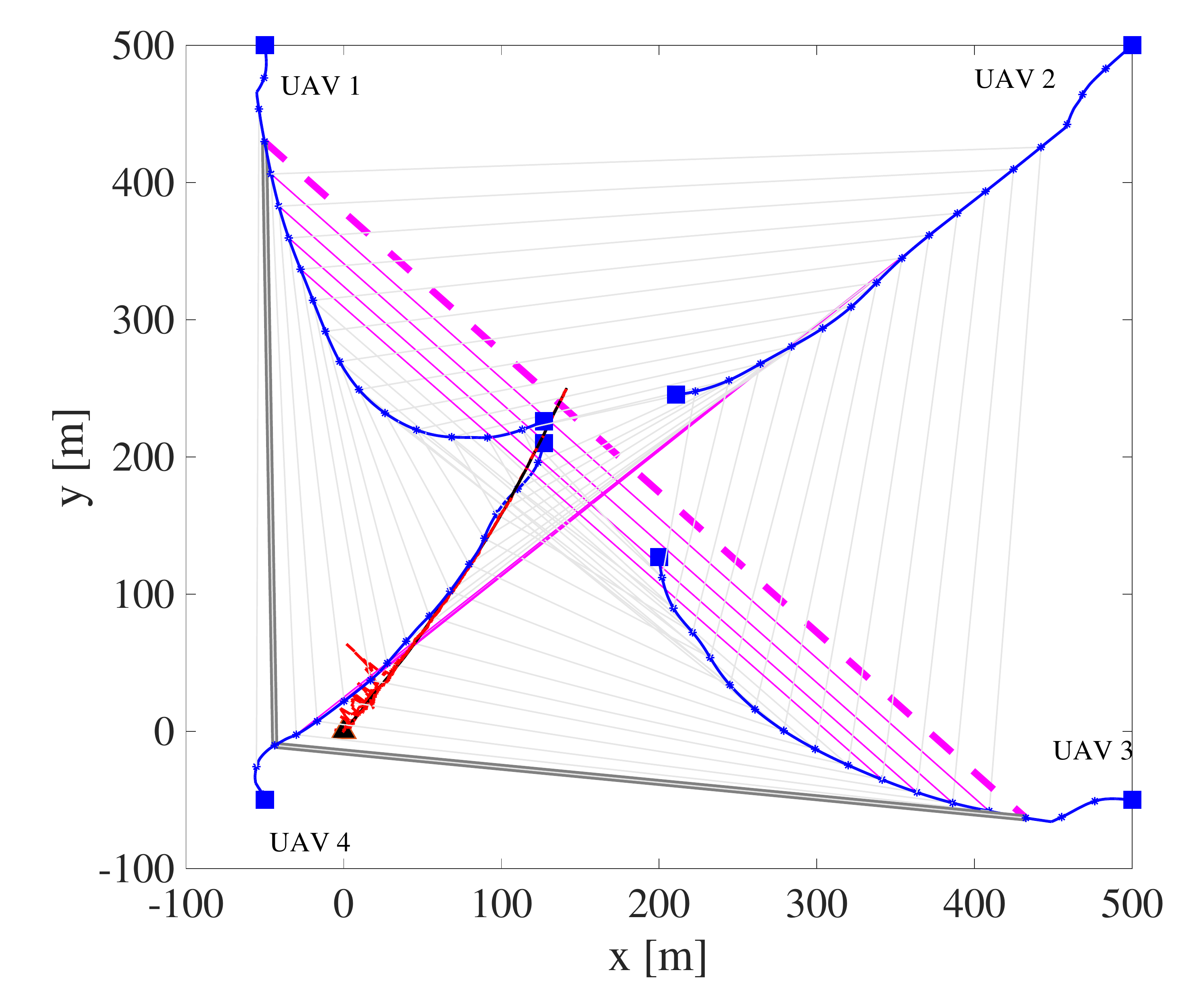} \includegraphics[width=0.3\linewidth,draft=false]{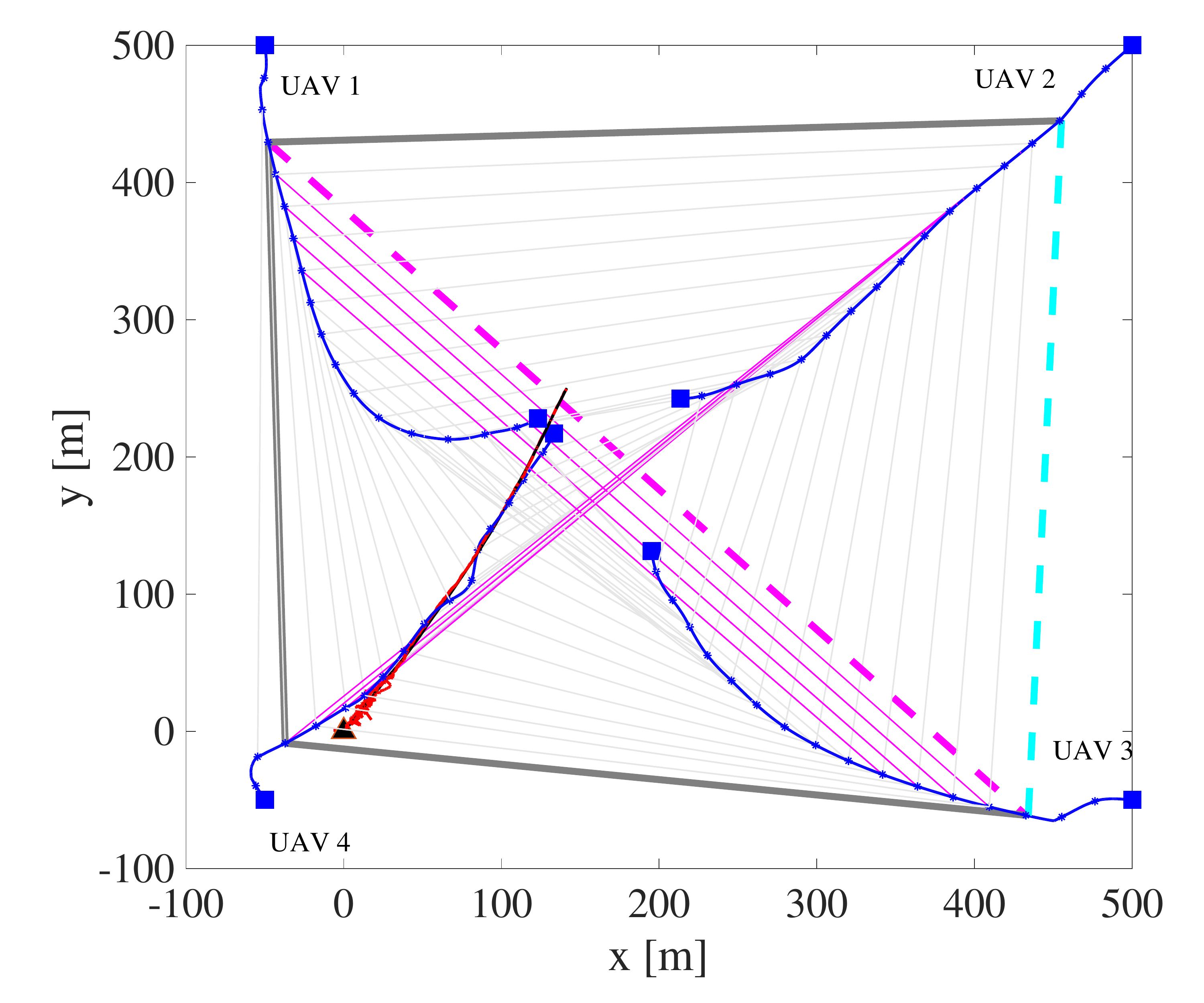}}
\centerline{
\includegraphics[width=0.45\linewidth,draft=false]{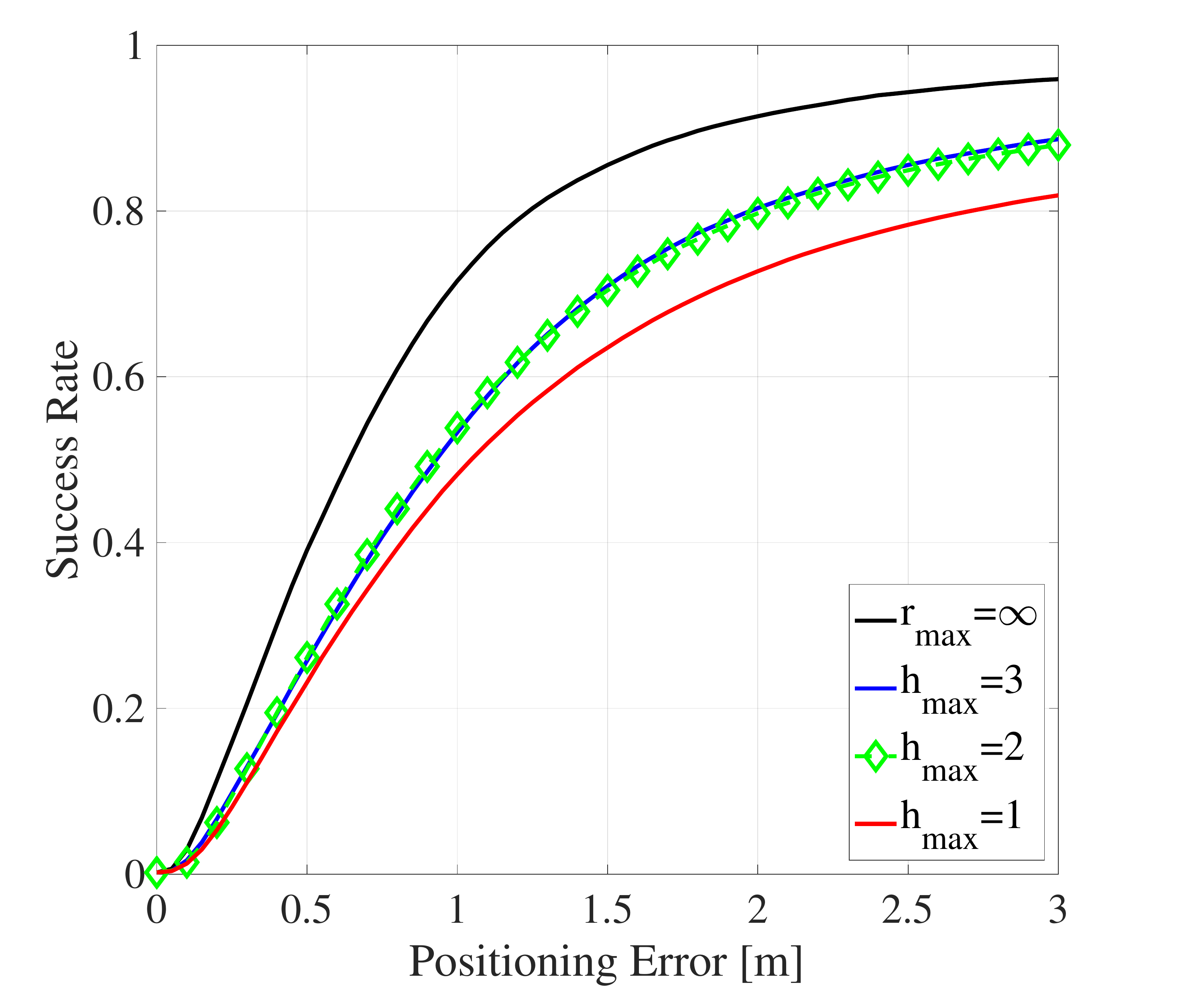}
}
\caption{Plot of the first $K=500$ instants of the trajectories with highlighted multi-hop connections. Top (from the left to the right): Single-hop scenario ($\hmax=1$); double-hop scenario ($\hmax=2$) with links reported with grey and magenta lines when a 1- or 2-hop is established, respectively; and triple-hop scenario ($\hmax=3$) with links reported with grey, magenta and cyan lines when a 1- or 2- or 3-hop is established, respectively. Bottom: Success rate as a function of the maximum number of hops.} 
\label{fig:multihop_sc}
\end{figure}

\begin{figure}[t]
\psfrag{k1}[rc][rc][0.5]{$k=1$}
\psfrag{k2}[lc][lc][0.5]{$500$}
\psfrag{k3}[lc][lc][0.5]{$1000$}
\psfrag{k4}[rc][rc][0.5]{$1500$}
\psfrag{k5}[lc][lc][0.5]{$2000$}
\psfrag{k6}[c][c][0.5]{$2500$}
\psfrag{k7}[lc][lc][0.5]{$3000$}
\psfrag{x}[c][c][0.5]{x [km]}
\psfrag{y}[c][c][0.5]{y [km]}
\psfrag{z}[c][c][0.7]{z [km]}
\psfrag{OB5movimento}[lc][lc][0.5]{Bearing-only, $4$ UAV-radars}
\psfrag{OB5terrestri}[lc][lc][0.5]{Bearing-only, $4$ terrestrial radars}
\psfrag{OR4movimento}[lc][lc][0.5]{Ranging-only, $4$ UAV-radars}
\psfrag{OR4terrestri}[lc][lc][0.5]{Ranging-only, $4$ terrestrial radars}
\psfrag{SINGOLONODOOOOOOO}[lc][lc][0.5]{Full sensing, $1$ terrestrial radar}
\psfrag{SuccessRate}[c][c][0.8]{Success Rate, $\mathsf{SR}\left( e_{\mathsf{th}} \right)$}
\psfrag{PositioningError}[c][c][0.8]{Positioning Error, $e_{\mathsf{th}}$, [m]}
\centerline{
\includegraphics[width=0.5\textwidth,draft=false]{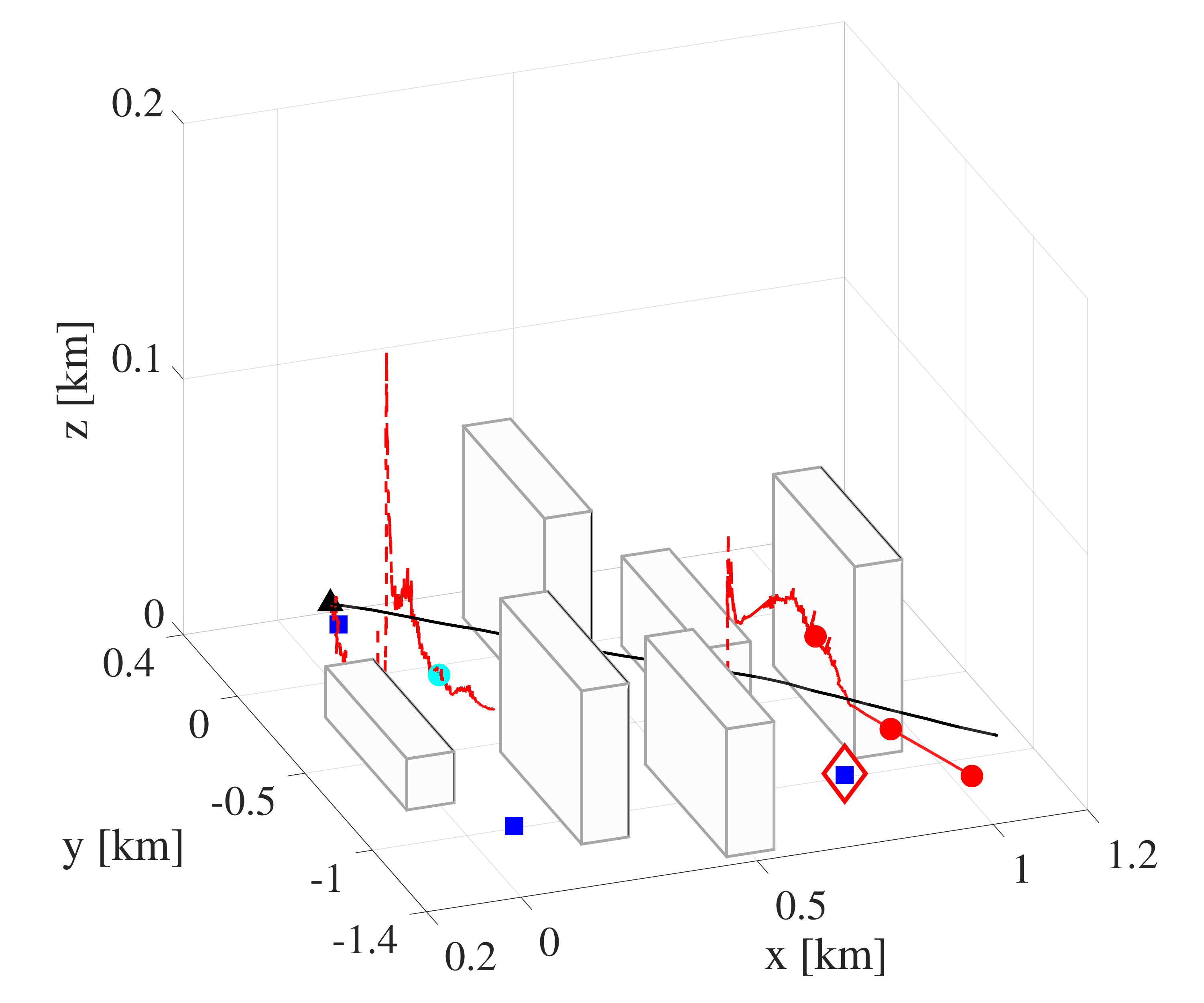} 
\includegraphics[width=0.5\textwidth,draft=false]{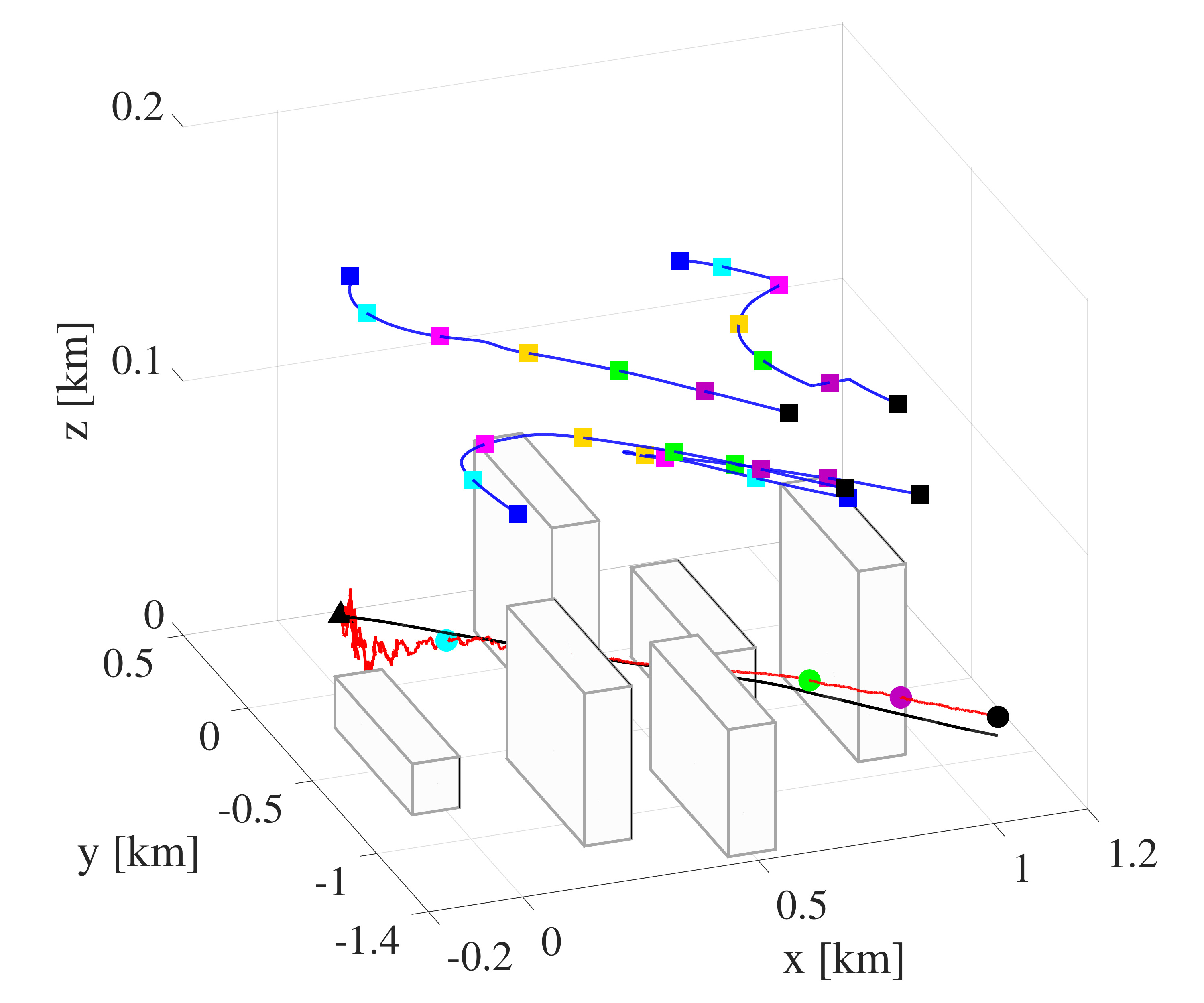} 
}
\vspace{0.3cm}
\centerline{\includegraphics[width=0.45\linewidth,draft=false]{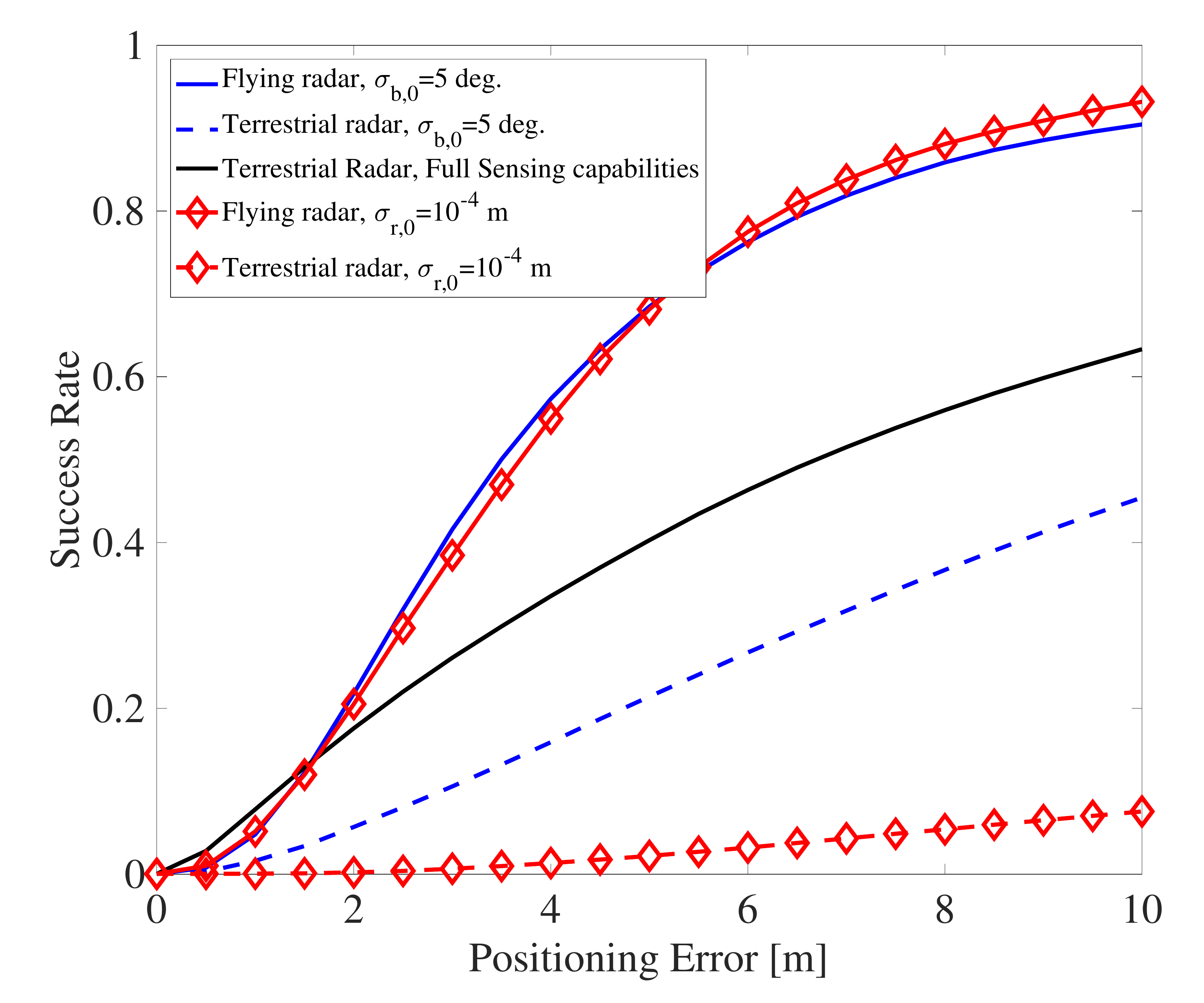}} 
\caption{Simulated scenarios in \ac{NLOS} conditions and success rate as a function of radar network configuration: comparison between terrestrial/fixed  and flying/dynamic radars, and success rate results (bottom).
} 
\label{fig:NLOSres}
\end{figure}

In this section, we analyze the performance of a \ac{DRN} in different conditions: by changing the number of \acp{UAV}; by varying their sensing capabilities;  by dealing with different \ac{RCS}; by varying the number of communication hops; and by operating in {LOS}-\ac{NLOS} channel conditions. 
The investigated scenarios are displayed in Figs.~\ref{fig:simulatedscenario_LOS}-\ref{fig:NLOSres}, with environments covering more than one square kilometer.
\begin{table}[t!]
\caption{RMSE [m] on target position for the configurations of Fig.\ref{fig:simulatedscenario_LOS}}
\label{tab:sim_RMSE}
  \centering
\begin{tabular}{lccc}
& $N=4$& $N=6$ & $N=10$    \\
\hline
\textit{Ranging-Only, $\srange=10^{-4}$} m &  0.53 & 0.33 & 0.15 \\
\textit{Ranging-Only, $\srange=10^{-2}$} m & 1.55 & 0.82 & 0.64 \\
\textit{Bearing-Only, $\sstb=5$} deg &  1.38 & 0.85 & 0.82 \\
\textit{Bearing-Only, $\sstb=20$}  deg& 3.85 & 2.27 & 1.55 \\
\end{tabular}
\end{table}
In the simulations, the target mobility in \eqref{eq:dyn_target} was modeled according to a random walk model \cite{dardari2015indoor}, with
\begin{align}\label{eq:straightline}
&\mathbf{A} \!=\! \left[\begin{array}{cc}
    \mathbf{I}_{3 \times 3} &  \Delta t \, \mathbf{I}_{3 \times 3} \\
    \mathbf{0}_{3 \times 3} & \mathbf{I}_{3 \times 3}
\end{array} \right],  \mathbf{Q} \!=\!\left[ \begin{array}{cc}
     \frac{\Delta t^3}{3} \, \Wok&  \frac{\Delta t^2}{2}\, \Wok  \\
     \frac{\Delta t^2}{2} \, \Wok &\Delta t \, \Wok
\end{array} \right],
\end{align}
where $\Wok=\text{diag} \left( \Woxk, \, \Woyk, \, \Wozk \right)=\left( 10^{-5}, \, 10^{-5}, \, 0 \right)$ is a diagonal matrix containing the variances of the process noise in each direction. The number of \acp{UAV} and the target \ac{RCS} were set to $6$ and $0.1\,$m$^2$, if not otherwise indicated. 
The safety distances, i.e., $\dmino$, $\dmint$ and $d_{\mathsf{O}}^*$, were all fixed at $5\,$m, the number of Monte Carlo iterations and the trajectory time steps at $100$ and $3000$ (each time step lasts 1 second), respectively.
A communication range of $900$ m between the \acp{UAV} and a single hop were considered \cite{wang2019autonomous,wang2019cooperative}, if not otherwise indicated. We initialized the \ac{EKF} as $\mathbf{m}_i^{(0)}=\mathbf{0}_{6 \times 1}$ and $\mathbf{P}_i^{(0)}= \text{diag}\left(20^2 \cdot \mathbf{I}_{3 \times 3},\, 0.5^2 \cdot \mathbf{I}_{3 \times 3}  \right)$.

To compare the results, the success rate was  evaluated as
\begin{align}
&\mathsf{SR}\left( e_{\mathsf{th}} \right)=  \frac{1}{K\,\Nmc\, N} \sum_{k=1}^{K}\, \sum_{i=1}^{N} \sum_{m=1}^{\Nmc} \mathbf{1}\left(e_{\mathsf{th}}-e_{im}^{(k)} \right),
\label{eq:successrate} 
\end{align}
where $\Nmc$ is the number of Monte Carlo iterations, $K$ is the number of time steps, $\mathbf{1}\left( x \right)$ is the unit step function that is equal to $1$ if $x \geq 0$ and $0$ otherwise,  $e_{im}^{(k)}$ is the estimation error of the target position at the $i$-th \ac{UAV} for the $m$-th Monte Carlo iteration, where  $e_{i,m}^{(k)}=\lVert \eposkm - \posk \rVert_2$, and $e_{\mathsf{th}}$ is a localization threshold.

In the simulations of Fig.~\ref{fig:simulatedscenario_LOS},
the initial positions of \acp{UAV} were at the vertexes of a square lying on the $XY$-plane with $x_i^{(0)}=\left[-50,\, -50,\, 500,\, 500\right]^{\mathsf{T}}\,$m, $y_i^{(0)}=\left[-50,\, 500,\, -50,\, 500\right]^{\mathsf{T}}\,$m, and $z_i^{(0)} \sim \mathcal{U}\left[80, 150 \right]\,$m, while the target initial position and velocity were $\left[0, 0, 90\right]^\mathsf{T}\,$m and $\left[-0.3, 0.4, 0\right]^\mathsf{T}\,$m/s.

In Fig.~\ref{fig:simulatedscenario_LOS}, we present qualitative examples of estimated \ac{UAV} trajectories for different sensing capabilities (ranging and bearing) and considering $N=4$. The trajectories of \acp{UAV} are reported as  blue lines and the positions are displayed with blue square markers for the initial and last time instants. The initial target position is drawn with a black triangle and its actual trajectory with a continuous black line. The estimated trajectory of the target is marked with a red dotted line. As can be seen, after an initial transient, the \acp{UAV} of the \ac{DRN} jointly surround the target.

Given this scenario, in Table~\ref{tab:sim_RMSE}, we show the tracking performance in terms of average \ac{RMSE} by varying the measurement accuracy and considering different number of \acp{UAV}. The \ac{RMSE} on position and velocity was averaged over the number of discrete time instants and over the number of \acp{UAV}. 
A group of four radars with only ranging capability and  accuracy of $\srange=10^{-2}\,$m  obtains approximately the same tracking performance of four radars with only bearing capability and accuracy of about $5^\circ\,$degrees. 
Instead, when considering a better performing radar, such as the \ac{FMCW} radar in \cite{Radar}  (i.e., with $\srange \approx 10^{-4}\,$m), the average localization accuracy is below $1\,$m. 

In Fig.~\ref{fig:results_SR_N}, we provide the success rate evaluated as in \eqref{eq:successrate} by varying the number of \acp{UAV} and the sensing capabilities. A localization error lower than $1\,$m can be achieved in nearly $80 \%$ of the cases with $N=4$ drones with either a reference ranging accuracy of $10^{-2}\,$m or a bearing accuracy of $5^\circ\,$degrees. This is also confirmed by  Fig.~\ref{fig:results_SR} where several ranging and bearing errors were tested.

We now investigate the impact of the Doppler shifts and target \ac{RCS} on the tracking performance with a fixed number of \acp{UAV} ($N=6$). 
In Fig.~\ref{fig:results_SR_RCS}-left, we show the success rate by considering ranging measurements and the presence of Doppler shifts with different chirp gains (i.e., $\Nchirp=64$-$256$) and different ranging accuracies. It can be observed that relying on Doppler shifts in addition to ranging measurements is beneficial especially when ranging is not sufficiently accurate: by fixing the desired localization error to $1\,$meter, the percentage increase experienced by adding Doppler shifts in the measurement vector is approximately of $100 \%$ with a ranging error of $\srange=10^{-1}\,$m (with $\Nchirp=256,\,\,\, \forall i=1, \ldots, N$) whereas there are no evident improvements for $\srange=10^{-3}\,$m. Finally, in Fig.~\ref{fig:results_SR_RCS}-right, we plot the success rate  as a function of the target \ac{RCS}. It is interesting to notice that a \ac{UAV} with a \ac{RCS} of $0.01\,$m can be localized in the $90 \%$ of cases with an error lower than $1$ meter provided that a sensor with a ranging accuracy of $10^{-3}\,$m is adopted. 

In Fig.~\ref{fig:multihop_sc}, we study the impact of a multi-hop exchange of measurements by limiting the number of temporal steps to $K=500$ because the impact of multi-hops is more evident at the beginning of the trajectory. The ranging accuracy was $\srange=10^{-3}\,$m, the number of \acp{UAV} to $N=4$, and a communication range of $\rmax=505\,$m was considered.  In Fig.~\ref{fig:multihop_sc}, from the left to the right, we have plotted the single hop scenario with links depicted with grey lines, the two-hop schenario, i.e., $\hmax=2$, with magenta lines and $3-$hop case with cyan lines, respectively. For example, in the single hop case, \ac{UAV}\,1 is only connected with \ac{UAV}\,4 at time instant $k=30$ because $\hmax=1$ and by having $\rmax=505\,$m only \ac{UAV}\,4 is in the neighboring set of \ac{UAV}\,1. Contrarily, when $\hmax=2$, it is also connected with \ac{UAV}\,3 through \ac{UAV}\,4. This means that the ranging information collected by \ac{UAV}\,3 will be available at \ac{UAV}\,1 after two time instants. Apart from an initial transient when the multi-hop propagation can be helpful as it allows to connect nodes otherwise unreachable, for the majority of the navigation time, a single-hop is sufficient thanks to the fact that the navigation control is conceived for minimizing the tracking error and, consequently, for  minimizing the \ac{UAV}-target and inter-\acp{UAV} distances. This is also confirmed from the results plotted in Fig.~\ref{fig:multihop_sc}-(bottom) in terms of success rate.
\begin{table}[t!]
\caption{RMSE for the configurations of Fig.\ref{fig:NLOSres}}
\label{tab:sim_ostacoli}
  \centering
\begin{tabular}{lcccc}
&\multicolumn{2}{c}{RMSE on position [m]}    & \multicolumn{2}{c}{RMSE on velocity  [m/s]}   \\
& Terr. Rad. & Flying Rad. & Terr. Rad. & Flying Rad.      \\
\hline
\textit{Ranging-Only}  & 65.17   & 5.07  & 0.12   & 0.05  \\
\textit{Bearing-Only}  &  17.43 & 5.70   &  0.071 &  0.063  
\end{tabular}
\end{table}

At this point, we aim at comparing the performance of a \ac{DRN} in presence of obstacles in order to assess the advantages of \acp{DRN} with respect to terrestrial fixed radar networks. To this purpose, we consider the scenario of Fig.~\ref{fig:NLOSres} where obstacles are depicted with grey parallelepipeds, the \acp{UAV} composing the \ac{DRN} with squared markers of different colors (every $500$ time steps), and the terrestrial radars with squared blue markers. The ranging and bearing errors were  $10^{-4}\,$m and $5^{\circ}\,$degrees, respectively.
In the \ac{DRN}, the \ac{UAV} initial positions were  $x_i^{(0)}=\left[100,\, 100,\, 800,\, 800 \right]^{\mathsf{T}}\,$m, $y_i^{(0)}=\left[-1000,\, 300,\, 300,\, -1000 \right]^{\mathsf{T}}\,$m, and with a \ac{UAV} height $z_i^{(0)} \sim \mathcal{U} \left[90, 150 \right]\,$m.  The target altitude was set to $30\,$m, and its trajectory followed the dynamics described by \eqref{eq:straightline}.
For a better comparison, two situations with a fixed deployment of radar sensors were considered: one with a single terrestrial radar with full sensing capabilities (capable of retrieving ranging, bearing and Doppler shift information) represented with a red diamond in Fig.~\ref{fig:NLOSres}-top, and another where, for fairness of comparison, the fixed radar network is with the same number ($N=4$) and sensing capabilities of \acp{UAV}. 
These radar configurations are compared in Fig.~\ref{fig:NLOSres}-bottom showing the superiority of a dynamic radar configuration over terrestrial networks in terms of success rate. Moreover, the \ac{RMSE} results on position and velocity are reported in Table II. In the case of a single terrestrial radar with full sensing capabilities, the \ac{RMSE} on position and velocities is of  $11.36\,$m and $0.06\,$m/s, respectively.

\section{Conclusion}
In this paper, the idea of a UAV dynamic radar network for the tracking of a non-cooperative (e.g., unauthorized) UAV  has been described.
In contrast with current on-ground radar systems, the UAV network provides new degrees of freedom thanks to its reconfigurability and flexibility. Moreover, the UAVs are considered autonomous in navigating and estimating their best trajectory to minimize the tracking error of the dynamic target. 
The proposed network has  heterogeneous sensing capabilities and estimates are shared among  the \acp{UAV}.  In this sense, the \ac{UAV} cooperation can significantly increase the tracking accuracy  without impacting the communication latency.  
The proposed control law aimed at minimizing an  information-driven cost function derived starting from measurements and estimates exchanged by the \acp{UAV} at each time instant. 

Results demonstrate that having a flexible network instead of a terrestrial deployment of radars helps in preventing NLOS conditions and, thus, in better tracking a non-cooperative target.
Moreover, even if the intruder is a small UAV (a target with \ac{RCS} of $0.1\,$m$^2$ or less), the positioning performance is below $1\,$m most of the time, provided that a radar sensor with a millimeter ranging accuracy is available on-board, as for example a \ac{FMCW} radar operating at $77\,$GHz. 
The same performance can be obtained with bearing  measurements given an angular accuracy of about $5^\circ\,$degrees.  Finally the use of Doppler shift estimates is beneficial to retrieve the velocity of the target instead of inferring it from position estimates. For this reason, the impact of the Doppler shift estimates is more valuable in the case where the ranging error is larger.
Future directions of research include the development of a control law able to maximize the expected information metric over a longer horizon (non-myopic approach) in order to deal better with complex and dynamic environments. 

\section*{Appendix A}
\label{sec:appA}

The elements in \eqref{eq:Gdhop}  provide the geometric matrix relative to Doppler shift measurements, and they  are given by
\begin{align}
\Gjldop\left( \posk, \posjl \right)&=\left[ \nabla_{\posk}^{\mathsf{T}}\left(\fdjl \right)\, \nabla_{\posk}\left(\fdjl \right) \right]_{\statek=\mathbf{s}_{0}^{-}}  = \left(\frac{\gamma}{2\, \lambda} \right)^2\,  \left[
    \begin{array}{ccc}
         \gxx & \gxy  & \gxz \\
         \gxy & \gyy  & \gyz \\
         \gxz & \gyz  & \gzz \\
    \end{array}
    \right],
\end{align}
with $\mathbf{s}_{0}^{-}=\statekpred$ and
\begin{align}
    &\gxx \!=\! \left(- \sph\, \sth \,  \omegaz + \cth \, \omegay\right)^2, \nonumber \\
    & \gyy \!=\! \left( \cph \, \sth \,  \omegaz {-} \cth \, \omegax \right)^2, \nonumber \\
    & \gzz \!=\! \left( -   \cph \, \sth\,  \omegay + \sph\, \sth\,  \omegax\right)^2, \nonumber \\
    &\gxy \!=\! \left(- \sph\, \sth \,  \omegaz + \cth \, \omegay \right)\!  \left( \cph \, \sth \,  \omegaz {-} \cth \, \omegax\right), \nonumber\\
    &\gxz \!=\! \left(\sph \sth \omegaz - \cth \omegay\right)\! \left( \cph  \sth  \omegay - \sph \sth \omegax\right), \nonumber\\
    &\gyz \!=\! \left( \cph  \sth   \omegaz {-} \cth  \omegax \right)\! \left( -   \cph  \sth  \omega_y + \sph \sth\,  \omegax\right). \nonumber
\end{align}
%

\section*{Appendix B}
\label{sec:appB}

In this appendix, the elements of the \textit{information matrix} in \eqref{eq:BIM2}, that is,
\begin{align}
\FIMik\left(\posk; \kposi\right)&=\left[
    \begin{array}{ccc}
         \Jxx & \Jxy  & \Jxz \\
         \Jxy & \Jyy  & \Jyz \\
         \Jxz & \Jyz  & \Jzz \\
    \end{array}
    \right],
\end{align}
are expanded in scalar notation. In particular, we have
\begin{align} 
\Jxx &=\Infikxxprev \!+\! \sum_{j=1}^{\lvert \Nneighi \rvert} \! \pjl\, \left\{ \frac{\flagrange\, \gamma^2}{4\, \left(\sjlr \right)^2} \, \left( \cph  \sth \right)^2  +   \frac{\flagbear}{\left(\sjlb \, \djlk \right)^2 } \, \left[\left(\frac{  \sph}{  \sth}\right)^2 +\left( \cph \, \cth\right)^2\right]   \right. \nonumber \\
& \left.   + \frac{\flagdop \, \gamma^2\, }{4\,\lambda^2\, \left( \sjldop \right)^2}\, \gxx \right\},  
\end{align}
\begin{align}
\Jxy &= \, \Infikxyprev + \sum_{j=1}^{\lvert \Nneighi \rvert}  \pjl  \left\{ \frac{ \flagrange\, \gamma^2}{4\, \left(\sjlr \right)^2} \, \sph\, \cph \, \left( \sth \right)^2  + \flagdop\,\frac{  \gamma^2\, \gxy}{4\,\lambda^2\, \left( \sjldop \right)^2} - \frac{\flagbear}{\left(\sjlb\, \djlk \right)^2 }   \frac{\sph\, \cph }{\left(\sth\right)^2}     \right. \nonumber \\
&\left. +\frac{\flagbear}{\left(\sjlb\, \djlk \right)^2 }   \sph \, \cph\, \left( \cth \right)^2 \right\},  \\
\Jxz&=\,\Infikxzprev + \sum_{j=1}^{\lvert \Nneighi \rvert}\, \pjl\, \left\{ \frac{\flagrange \, \gamma^2}{4\, \left(\sjlr \right)^2}\, \cph \, \sth \, \cth  - \frac{ \flagbear}{\left(\sjlb \, \djlk\right)^2} \, \cph \, \sth\, \cth    +  \frac{\flagdop\, \gamma^2}{{2}\,\lambda^2\, \left( \sjldop \right)^2}\, \gxz    \right\},  \\
\Jyy&=  \Infikyyprev + \sum_{j=1}^{\lvert \Nneighi \rvert} \pjl\,   \left\{  \frac{\flagrange\, \gamma^2}{4\, \left(\sjlr \right)^2} \,\left( \sph\,  \sth \right)^2 + \frac{\flagbear}{\left(\sjlb\, \djlk \right)^2} \left[ \left(\frac{\cph}{\sth} \right)^2+  \left( \sph \, \cth \right)^2  \right]    \right.  \nonumber \\
&\left. + \flagdop\, \frac{\gamma^2}{4\,\lambda^2\, \left( \sjldop \right)^2}\,\, \gyy\right\},  \\
 \Jzz&= \, \Infikzzprev + \sum_{j=1}^{\lvert \Nneighi \rvert}  \pjl  \left\{  \frac{\flagrange\, \gamma^2}{4\, \left(\sjlr \right)^2} \, \left( \cth \right)^2 +  \frac{\flagdop\, \gamma^2}{4\,\lambda^2\, \left( \sjldop \right)^2}\, \gzz + \flagbear\,\frac{1}{\left(\sjlb \djlk \right)^2}  \, \left( \sth \right)^2  \right\},  \\
\Jyz &= \, \,\Infikyzprev + \sum_{j=1}^{\lvert \Nneighi \rvert} \pjl \left\{  \frac{\flagrange\, \gamma^2}{4\, \left(\sjlr \right)^2} \, \sth\, \cth\, \sph + \frac{\flagdop\, \gamma^2}{4\,\lambda^2\, \left( \sjldop \right)^2}\, \gyz \,  -  \, \frac{\flagbear}{\left(\sjlb \, \djlk \right)^2}  \, \sph\, \sth\, \cth  \right\}. \label{eq:jyz}
\end{align}

\section*{Appendix C}
\label{sec:appC}

In this appendix, we derive the analytical expressions for control signals in \eqref{eq:update_g2}. More specifically, we determine the term $\nabla_{\posi} \, \mathcal{C}\left( \eposk, \kposi\right)$. According to \eqref{eq:costfunction}, we have
\begin{align}
\nabla_{\posi} \, \mathcal{C}\left( \eposk, \kposi\right)
&= -\frac{\nablapi \left(\Dik \left( \kposi \right) \right)}{\Dik \left( \kposi \right)}, 
\end{align}
where $\Dik \left( \kposi \right)=\det \left(\FIMik\left( \eposk; \kposi  \right)\right) = \Jxx\, \Cxx + \Jxy\, \Cyx + \Jxz\, \Czx$ is the determinant of the {information matrix}, and $\Cxx=\Jyy \Jzz- \Jzy^2$  $\Cxy=\Cyx=\Jxz \Jzy- \Jxy\, \Jzz$, $\Cxz=\Czx=\Jxy \Jyz- \Jxz\, \Jyy$, $\Cyy=\Jxx\, \Jzz- \Jxz^2$, $\Cyz=\Jxz\, \Jyx- \Jxx\, \Jyz$, $\Czz=\Jxx\, \Jyy- \Jxy^2$ are the cofactors of the inverse.
Consequently, the derivatives of the cost functions are
\begin{align}\label{eq:d1}
\nablapi \left(\Dik \left( \kposi \right) \right)&= \left( \nablapi \Jxx \right) \Cxx   +\left( \nablapi \Cxx \right) \Jxx+\left( \nablapi \Jxy \right) \Cyx+\left( \nablapi \Cyx \right) \Jxy  \nonumber \\
&+\left( \nablapi \Jxz \right)\Czx+\left( \nablapi \Cxz \right) \Jzx\,, 
\end{align}
where
\begin{align}
\nablapi \Cxx &= \left( \nablapi \Jyy \right) \Jzz +  \left( \nablapi \Jzz\right) \Jyy - 2\, \Jzy\, \left( \nablapi \Jzy\right), 
\nonumber \\
\nablapi \Cyx &=  \left( \nablapi  \Jxz \right) \Jzy +\Jxz \left( \nablapi \Jzy\right) - \left( \nablapi \Jxy\right)\, \Jzz - \Jxy\, \left( \nablapi  \Jzz\right), \nonumber
\end{align}
\begin{align}\label{eq:d2}
\nablapi \Cxz &= \left( \nablapi \Jxy \right) \Jyz + \Jxy \left( \nablapi  \Jyz \right) - \left( \nablapi  \Jxz \right) \, \Jyy - \Jxz\, \left( \nablapi  \Jyy \right) .
\end{align}
Starting from \eqref{eq:jyz}, it is straightforward to derive \eqref{eq:d2}. 

\bibliographystyle{IEEEtran}

\begin{thebibliography}{10}
\providecommand{\url}[1]{#1}
\csname url@samestyle\endcsname
\providecommand{\newblock}{\relax}
\providecommand{\bibinfo}[2]{#2}
\providecommand{\BIBentrySTDinterwordspacing}{\spaceskip=0pt\relax}
\providecommand{\BIBentryALTinterwordstretchfactor}{4}
\providecommand{\BIBentryALTinterwordspacing}{\spaceskip=\fontdimen2\font plus
\BIBentryALTinterwordstretchfactor\fontdimen3\font minus
  \fontdimen4\font\relax}
\providecommand{\BIBforeignlanguage}[2]{{%
\expandafter\ifx\csname l@#1\endcsname\relax
\typeout{** WARNING: IEEEtran.bst: No hyphenation pattern has been}%
\typeout{** loaded for the language `#1'. Using the pattern for}%
\typeout{** the default language instead.}%
\else
\language=\csname l@#1\endcsname
\fi
#2}}
\providecommand{\BIBdecl}{\relax}
\BIBdecl

\bibitem{zhao2019uav}
N.~Zhao \emph{et~al.}, ``{UAV}-assisted emergency networks in disasters,''
  \emph{IEEE Wireless Commun.}, vol.~26, no.~1, pp. 45--51, 2019.

\bibitem{shakeri2019design}
R.~Shakeri \emph{et~al.}, ``Design challenges of multi-{UAV} systems in
  cyber-physical applications: A comprehensive survey, and future directions,''
  \emph{IEEE Commun. Surveys \& Tutorials}, 2019.

\bibitem{mozaffari2018beyond}
M.~Mozaffari, A.~T.~Z. Kasgari, W.~Saad, M.~Bennis, and M.~Debbah, ``Beyond
  {5G} with {UAV}s: Foundations of a {3D} wireless cellular network,''
  \emph{IEEE Trans. Wireless Commun.}, vol.~18, no.~1, pp. 357--372, 2018.

\bibitem{gangula2017trajectory}
R.~Gangula, P.~de~Kerret, O.~Esrafilian, and D.~Gesbert, ``Trajectory
  optimization for mobile access point,'' in \emph{Proc. 2017 51st Asilomar
  Conf. Signals, Sys., Comput.}\hskip 1em plus 0.5em minus 0.4em\relax IEEE,
  2017, pp. 1412--1416.

\bibitem{chen2017learning}
J.~Chen, U.~Yatnalli, and D.~Gesbert, ``Learning radio maps for {UAV}-aided
  wireless networks: A segmented regression approach,'' in \emph{Proc. 2017
  IEEE Int. Conf. Commun. (ICC)}.\hskip 1em plus 0.5em minus 0.4em\relax IEEE,
  2017, pp. 1--6.

\bibitem{cerwall2015ericsson}
P.~Cerwall, P.~Jonsson, R.~M{\"o}ller, S.~B{\"a}vertoft, S.~Carson, and
  I.~Godor, ``Ericsson mobility report,'' \emph{On the Pulse of the Networked
  Society. Hg. v. Ericsson}, 2015.

\bibitem{guvenc2018detection}
I.~Guvenc, F.~Koohifar, S.~Singh, M.~L. Sichitiu, and D.~Matolak, ``Detection,
  tracking, and interdiction for amateur drones,'' \emph{IEEE Commun. Mag.},
  vol.~56, no.~4, pp. 75--81, 2018.

\bibitem{koohifar2018autonomous}
F.~Koohifar, I.~Guvenc, and M.~L. Sichitiu, ``Autonomous tracking of
  intermittent {RF} source using a {UAV} swarm,'' \emph{IEEE Access}, vol.~6,
  pp. 15\,884--15\,897, 2018.

\bibitem{bisio2018unauthorized}
I.~Bisio, C.~Garibotto, F.~Lavagetto, A.~Sciarrone, and S.~Zappatore,
  ``Unauthorized amateur {UAV} detection based on {WiFi} statistical
  fingerprint analysis,'' \emph{IEEE Commun. Mag.}, vol.~56, no.~4, pp.
  106--111, 2018.

\bibitem{hugler2018radar}
P.~H{\"u}gler, F.~Roos, M.~Schartel, M.~Geiger, and C.~Waldschmidt, ``Radar
  taking off: New capabilities for {UAV}s,'' \emph{IEEE Microw. Mag.}, vol.~19,
  no.~7, pp. 43--53, 2018.

\bibitem{ezuma2019micro}
M.~Ezuma, O.~Ozdemir, C.~K. Anjinappa, W.~A. Gulzar, and I.~Guvenc,
  ``Micro-{UAV} detection with a low-grazing angle millimeter wave radar,''
  \emph{arXiv preprint arXiv:1902.05483}, 2019.

\bibitem{casbeer2006connectivity}
D.~Casbeer, A.~L. Swindlehurst, and R.~Beard, ``Connectivity in a {UAV}
  multi-static radar network,'' in \emph{AIAA Guidance, Navig., Control Conf.
  Exhibit}, 2006, p. 6209.

\bibitem{solomitckii2018technologies}
D.~Solomitckii, M.~Gapeyenko, V.~Semkin, S.~Andreev, and Y.~Koucheryavy,
  ``Technologies for efficient amateur drone detection in {5G} millimeter-wave
  cellular infrastructure,'' \emph{IEEE Commun. Mag.}, vol.~56, no.~1, pp.
  43--50, 2018.

\bibitem{paul2015extending}
B.~Paul and D.~W. Bliss, ``Extending joint radar-communications bounds for
  {FMCW} radar with doppler estimation,'' in \emph{Proc. 2015 IEEE Radar Conf.
  (RadarCon)}.\hskip 1em plus 0.5em minus 0.4em\relax IEEE, 2015, pp.
  0089--0094.

\bibitem{schuster2006performance}
S.~Schuster, S.~Scheiblhofer, L.~Reindl, and A.~Stelzer, ``Performance
  evaluation of algorithms for {SAW}-based temperature measurement,''
  \emph{IEEE Trans. Ultrason. Ferroelectr. Freq. Control}, vol.~53, no.~6, pp.
  1177--1185, 2006.

\bibitem{liu2019relative}
Y.~Liu, W.~Li, Q.~Lu, J.~Wang, and Y.~Shen, ``Relative localization of ground
  vehicles using non-terrestrial networks,'' in \emph{2019 IEEE/CIC Int. Conf.
  Commun. Workshops China (ICCC Workshops)}.\hskip 1em plus 0.5em minus
  0.4em\relax IEEE, 2019, pp. 93--97.

\bibitem{guerra2018collaborative}
A.~Guerra, N.~Sparnacci, D.~Dardari, and P.~M. Djuri{\'c}, ``Collaborative
  target-localization and information-based control in networks of {UAV}s,'' in
  \emph{Proc. 2018 IEEE 19th Inter. Workshop Signal Process. Adv. Wireless
  Commun. (SPAWC)}.\hskip 1em plus 0.5em minus 0.4em\relax IEEE, 2018, pp.
  1--5.

\bibitem{guerra2018joint}
A.~Guerra, D.~Dardari, and P.~M. Djuri{\'c}, ``Joint indoor localization and
  navigation of uavs for network formation control,'' in \emph{Proc. 2018 52nd
  Asilomar Conf. Signals, Sys., Comput.}\hskip 1em plus 0.5em minus 0.4em\relax
  IEEE, 2018, pp. 13--19.

\bibitem{guerra2019non}
A.~Guerra, D.~Dardari, and P.~M. Djuric, ``Non-centralized navigation for
  source localization by cooperative uavs,'' \emph{arXiv preprint
  arXiv:1910.12780}, 2019.

\bibitem{feger200977}
R.~Feger, C.~Wagner, S.~Schuster, S.~Scheiblhofer, H.~Jager, and A.~Stelzer,
  ``A {77-GHz FMCW MIMO radar based on an SiGe single-chip transceiver},''
  \emph{IEEE Trans. Microw. Theory and Techn.}, vol.~57, no.~5, pp. 1020--1035,
  2009.

\bibitem{folster2005automotive}
F.~Folster, H.~Rohling, and U.~Lubbert, ``An automotive radar network based on
  {77 GHz FMCW sensors},'' in \emph{Proc. IEEE Int. Radar Conf., 2005.}\hskip
  1em plus 0.5em minus 0.4em\relax IEEE, 2005, pp. 871--876.

\bibitem{hoffmann2016micro}
F.~Hoffmann, M.~Ritchie, F.~Fioranelli, A.~Charlish, and H.~Griffiths,
  ``Micro-doppler based detection and tracking of {UAV}s with multistatic
  radar,'' in \emph{Proc. 2016 IEEE Radar Conf. (RadarConf)}.\hskip 1em plus
  0.5em minus 0.4em\relax IEEE, 2016, pp. 1--6.

\bibitem{martinez2006optimal}
S.~Mart{\'i}nez and F.~Bullo, ``Optimal sensor placement and motion
  coordination for target tracking,'' \emph{Automatica}, vol.~42, no.~4, pp.
  661--668, 2006.

\bibitem{ragi2013uav}
S.~Ragi and E.~K. Chong, ``{UAV} path planning in a dynamic environment via
  partially observable {M}arkov decision process,'' \emph{IEEE Trans. Aerosp.
  Electron. Syst.}, vol.~49, no.~4, pp. 2397--2412, 2013.

\bibitem{kassas2015receding}
Z.~M. Kassas and T.~E. Humphreys, ``Receding horizon trajectory optimization in
  opportunistic navigation environments,'' \emph{IEEE Trans. Aerosp. Electron.
  Syst.}, vol.~51, no.~2, pp. 866--877, 2015.

\bibitem{dogancay2012uav}
K.~Dogancay, ``{UAV} path planning for passive emitter localization,''
  \emph{IEEE Trans. Aerosp. Electron. Syst.}, vol.~48, no.~2, pp. 1150--1166,
  2012.

\bibitem{wang2019autonomous}
C.~Wang, J.~Wang, Y.~Shen, and X.~Zhang, ``Autonomous navigation of {UAV}s in
  large-scale complex environments: A deep reinforcement learning approach,''
  \emph{IEEE Trans. Veh. Technol.}, vol.~68, no.~3, pp. 2124--2136, 2019.

\bibitem{cai2019integrated}
Y.~Cai and Y.~Shen, ``An integrated localization and control framework for
  multi-agent formation,'' \emph{IEEE Trans. Signal Process.}, vol.~67, no.~7,
  pp. 1941--1956, 2019.

\bibitem{opromolla2019airborne}
R.~Opromolla, G.~Inchingolo, and G.~Fasano, ``Airborne visual detection and
  tracking of cooperative {UAV}s exploiting deep learning,'' \emph{Sensors},
  vol.~19, no.~19, p. 4332, 2019.

\bibitem{ucinski2004optimal}
D.~Ucinski, \emph{Optimal measurement methods for distributed parameter system
  identification}.\hskip 1em plus 0.5em minus 0.4em\relax CRC Press, 2004.

\bibitem{meyer2015distributed}
F.~Meyer, H.~Wymeersch, M.~Fr{\"o}hle, and F.~Hlawatsch, ``Distributed
  estimation with information-seeking control in agent networks,'' \emph{IEEE
  J. Sel. Areas Commun.}, vol.~33, no.~11, pp. 2439--2456, 2015.

\bibitem{meyer2017scalable}
F.~Meyer, P.~Braca, P.~Willett, and F.~Hlawatsch, ``A scalable algorithm for
  tracking an unknown number of targets using multiple sensors,'' \emph{IEEE
  Trans. Signal Process.}, vol.~65, no.~13, pp. 3478--3493, 2017.

\bibitem{meyer2015distributedob}
F.~Meyer, O.~Hlinka, H.~Wymeersch, E.~Riegler, and F.~Hlawatsch, ``Distributed
  localization and tracking of mobile networks including noncooperative
  objects,'' \emph{IEEE Trans. Signal Inf. Process. Netw.}, vol.~2, no.~1, pp.
  57--71, 2015.

\bibitem{tang2018autonomous}
S.~Tang and V.~Kumar, ``Autonomous flight,'' \emph{Annual Review of Control,
  Robotics, and Autonomous Systems}, vol.~1, pp. 29--52, 2018.

\bibitem{dardari2015indoor}
D.~Dardari, P.~Closas, and P.~M. Djuri{\'c}, ``Indoor tracking: Theory,
  methods, and technologies,'' \emph{IEEE Trans. Veh. Technol.}, vol.~64,
  no.~4, pp. 1263--1278, 2015.

\bibitem{isaacs2014quadrotor}
J.~T. Isaacs, F.~Quitin, L.~R.~G. Carrillo, U.~Madhow, and J.~P. Hespanha,
  ``Quadrotor control for {RF} source localization and tracking,'' in
  \emph{Proc. 2014 Int. Conf. {U}nmanned {A}ircraft {S}ystems (ICUAS)}.\hskip
  1em plus 0.5em minus 0.4em\relax IEEE, 2014, pp. 244--252.

\bibitem{xu20183d}
S.~Xu, K.~Do{\u{g}}an{\c{c}}ay, and H.~Hmam, ``{3D AOA} target tracking using
  distributed sensors with multi-hop information sharing,'' \emph{Signal
  Process.}, vol. 144, pp. 192--200, 2018.

\bibitem{sarkka2013bayesian}
S.~S{\"a}rkk{\"a}, \emph{Bayesian filtering and smoothing}.\hskip 1em plus
  0.5em minus 0.4em\relax Cambridge University Press, 2013, vol.~3.

\bibitem{dedecius2016sequential}
K.~Dedecius and P.~M. Djuri{\'c}, ``Sequential estimation and diffusion of
  information over networks: A bayesian approach with exponential family of
  distributions,'' \emph{IEEE Trans. Signal Process.}, vol.~65, no.~7, pp.
  1795--1809, 2016.

\bibitem{ivashko2013performance}
I.~Ivashko, O.~Krasnov, and A.~Yarovoy, ``Performance analysis of multisite
  radar systems,'' in \emph{Proc. 2013 European Radar Conf.}\hskip 1em plus
  0.5em minus 0.4em\relax IEEE, 2013, pp. 459--462.

\bibitem{ivashko2015topology}
------, ``Topology optimization of monostatic radar networks with wide-beam
  antennas,'' in \emph{Proc. 2015 European Radar Conf.)}.\hskip 1em plus 0.5em
  minus 0.4em\relax IEEE, 2015, pp. 133--136.

\bibitem{petitjean2018pimrc}
M.~Petitjean, S.~Mezhoud, and F.~Quitin, ``Fast localization of ground-based
  mobile terminals with a transceiver-equipped {UAV},'' in \emph{Proc. 2018
  29th Annual Int. Symp. Personal, Indoor, Mobile Radio Commun. (PIMRC)}.\hskip
  1em plus 0.5em minus 0.4em\relax IEEE, 2018.

\bibitem{bar2004estimation}
Y.~Bar-Shalom, X.~R. Li, and T.~Kirubarajan, \emph{Estimation with applications
  to tracking and navigation: theory algorithms and software}.\hskip 1em plus
  0.5em minus 0.4em\relax John Wiley \& Sons, 2004.

\bibitem{dardari2012satellite}
D.~Dardari, E.~Falletti, and M.~Luise, \emph{Satellite and terrestrial radio
  positioning techniques: a signal processing perspective}.\hskip 1em plus
  0.5em minus 0.4em\relax Elsevier, 2012.

\bibitem{luenberger1984linear}
D.~G. Luenberger, Y.~Ye \emph{et~al.}, \emph{Linear and nonlinear
  programming}.\hskip 1em plus 0.5em minus 0.4em\relax Springer, 1984, vol.~2.

\bibitem{bertsekas1995dynamic}
D.~P. Bertsekas, \emph{Dynamic programming and optimal control}.\hskip 1em plus
  0.5em minus 0.4em\relax Athena scientific Belmont, MA, 1995, vol.~1, no.~2.

\bibitem{fink2013robust}
J.~Fink, A.~Ribeiro, and V.~Kumar, ``Robust control of mobility and
  communications in autonomous robot teams,'' \emph{IEEE Access}, vol.~1, pp.
  290--309, 2013.

\bibitem{wang2019cooperative}
Y.~Wang, Y.~Wu, and Y.~Shen, ``Cooperative tracking by multi-agent systems
  using signals of opportunity,'' \emph{IEEE Trans. Commun.}, 2019.

\bibitem{Radar}
``{IWR1443 single-chip 76-GHz to 81-GHz mmWave sensor},''
  \url{http://www.ti.com/lit/wp/spyy005/spyy005.pdf}.

\end{thebibliography}

\end{document}